\documentclass[conference]{IEEEtran}
\usepackage{cite}

\usepackage{mathptmx} 
\usepackage{fancyhdr}
\usepackage[normalem]{ulem}
\usepackage[hyphens]{url}
\usepackage[keeplastbox]{flushend}

\usepackage{amsmath,amssymb,amsfonts} 
\usepackage{algorithmic}
\usepackage{graphicx}
\usepackage{textcomp}
\usepackage{xcolor}
\usepackage{multirow}

\definecolor{remove}{RGB}{0, 111, 111}
\definecolor{general}{RGB}{0, 102, 0}
\definecolor{r1}{RGB}{0, 102, 0}
\definecolor{r2}{RGB}{0, 102, 0}
\definecolor{r3}{RGB}{0, 102, 0}
\definecolor{r4}{RGB}{0, 102, 0}
\definecolor{r5}{RGB}{0, 102, 0}

\def\BibTeX{{\rm B\kern-.05em{\sc i\kern-.025em b}\kern-.08em
    T\kern-.1667em\lower.7ex\hbox{E}\kern-.125emX}}    

\begin{document}

\title{AWB-GCN: A Graph Convolutional Network Accelerator with Runtime Workload Rebalancing\vspace{0.05truein}}

\author{\Large  \vspace{0.02truein} Tong Geng$^{1, 2}$, Ang Li$^{2}$, Runbin Shi$^{4}$, Chunshu Wu$^{1}$, Tianqi Wang$^{1}$, Yanfei Li$^{5}$, \\ \vspace{0.16truein} \Large Pouya Haghi$^{1}$, Antonino Tumeo$^{2}$, Shuai Che$^{3}$, Steve Reinhardt$^{3}$, Martin C. Herbordt$^{1}$\\ \large$^{1}$Boston University \\ $^{2}$Pacific Northwest National Laboratory \\ \large$^{3}$Microsoft \\  $^{4}$The University of Hong Kong \\  $^{5}$Zhejiang University}


\maketitle
\pagestyle{plain}
\begin{abstract}
    Deep learning systems have been successfully applied to Euclidean data such as images, video, and audio. In many applications, however, information and their relationships are better expressed with graphs. Graph Convolutional Networks (GCNs) appear to be a promising approach to efficiently learn from graph data structures, having shown advantages in many critical applications.
    As with other deep learning modalities, hardware acceleration is critical. The challenge is that real-world graphs are often extremely large and unbalanced; this poses significant performance demands and design challenges.
    
   In this paper, we propose Autotuning-Workload-Balancing GCN (AWB-GCN) to accelerate GCN inference. To address the issue of workload imbalance in processing real-world graphs, three hardware-based autotuning techniques are proposed: dynamic distribution smoothing, remote switching, and row remapping. In particular, AWB-GCN continuously monitors the sparse graph pattern, dynamically adjusts the workload distribution among a large number of processing elements (up to 4K PEs), and, after converging, reuses the ideal configuration. Evaluation is performed using an Intel D5005 FPGA with five commonly-used datasets. Results show that 4K-PE AWB-GCN can significantly elevate PE utilization by 7.7$\times$ on average and demonstrate considerable performance speedups over CPUs (3255$\times$), GPUs (80.3$\times$), and a prior GCN accelerator (5.1$\times$).

\end{abstract}


\section{Introduction}

Deep learning paradigms such as Convolutional Neural Networks (CNNs) \cite{krizhevsky2012imagenet} and Recurrent Neural Networks (RNNs) \cite{mikolov2010recurrent} have been applied to a wide range of applications including image classification, video processing, speech recognition, and natural language processing \cite{kwon2019understanding,kwon2020maestro,geng19,geng18,Li19}. These paradigms, however, are only able to extract and analyze latent information from euclidean data such as images, video, audio, and text \cite{wu2020comprehensive, wang20, geng21}. As a result, the adoption of neural networks is greatly limited in fields with complex relationships among objects. A large (and increasing) number of applications use non-Euclidean data structures that are modeled as graphs.
Nodes and edges represent objects and relationships between those objects, respectively, as appropriate for the application. Most of these graphs have a tremendously large numbers of nodes; moreover, the node degree generally varies dramatically, often following a power law distribution \cite{gonzalez2012powergraph, abou2006multilevel, latapy2008main, xie2014distributed, aiello2001random, chung2004spectra, adamic2001search}. 


The irregularity of the graph data makes most of the existing Neural Network (NN) algorithms ill-suited;
critical feature extraction operations, such as convolutions, are no longer applicable. To tackle this issue, Graph Neural Networks (GNNs) have been proposed, in various forms, to extend deep learning approaches to graph data \cite{gori2005new,micheli2009neural,scarselli2008graph,li2015gated,dai2018learning,you2018graphrnn,abu2018watch,wu2020comprehensive,gao2018large}. Among various GNNs, the \emph{Graph Convolutional Network} (GCN), an approach that marries some ideas of CNNs to the distinct needs of graph data processing, has demonstrated significant potential and become one of the most important topics in NN-based graph research \cite{henaff2015deep,bruna2013spectral,defferrard2016convolutional,kipf2016semi, yun2019graph}.

With the rapid development of GCNs, designing dedicated hardware accelerators has become an urgent issue \cite{yan2020hygcn}. GCNs have already been investigated in a large number of real-world applications \cite{wu2020comprehensive}, including electric grid cascading failure analysis \cite{liu2020guiding}, prediction of chemical reactivity \cite{coley2019graph}, prediction of synthesized material properties \cite{xie2018crystal}, polypharmacy side-effect modeling \cite{zitnik2018modeling}, accurate advertisement in E-commerce \cite{yang2019aligraph}, and cybersecurity \cite{nguyen2018iot}. Many of these applications pose stringent constraints on latency and throughput.

\begin{figure}[t] 
\centering
\includegraphics[width=3.5in]{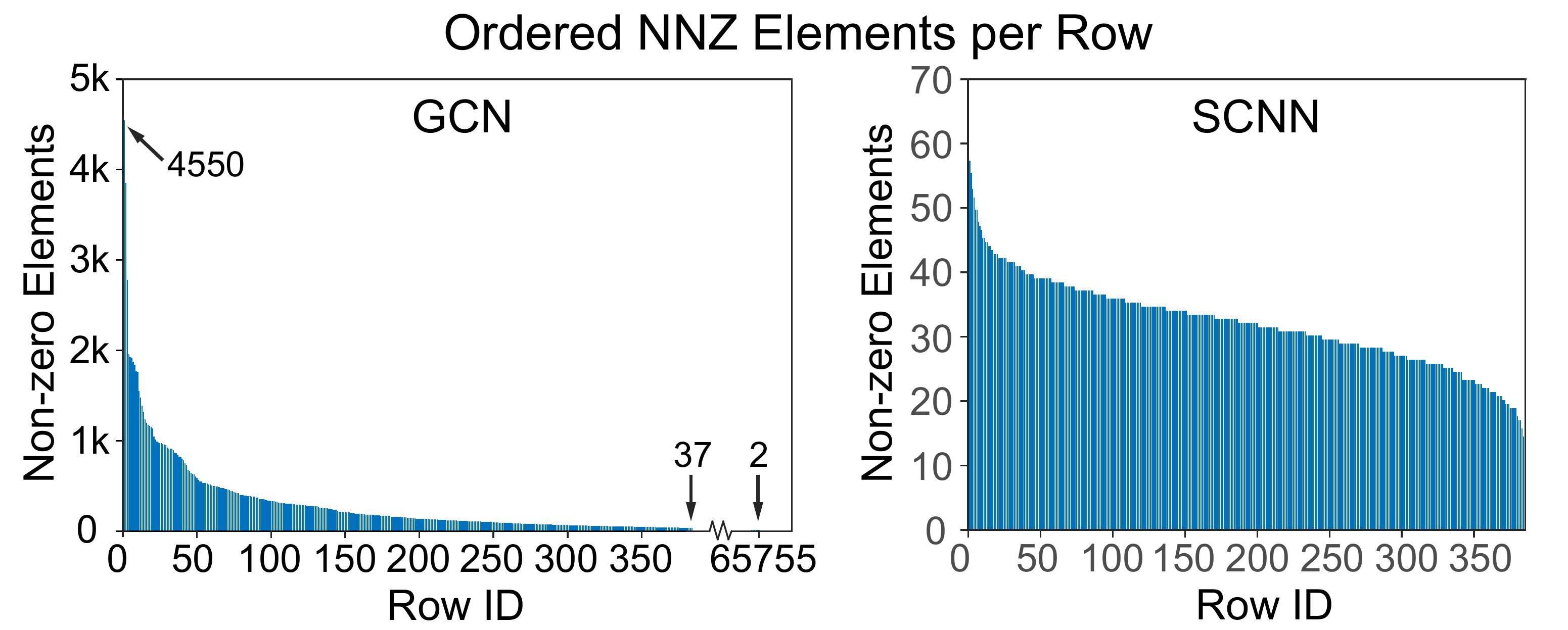}
\caption{Histograms show ordered non-zero per-row density. Left: Adjacency matrix of the NELL graph (avg. density: 0.0073\%) has most of non-zeros clustered in 70/66k rows. Right: Unstructured compressed AlexNet weight matrix (avg. density: 27\%) has workload roughly balanced across 384 rows.}
\label{power_law}
\end{figure}

Accelerators developed for other domains, such as the sparse-CNN-accelerator (SCNN) 
\cite{han2016eie, zhang2016cambricon, kim2017novel}, are not likely to be optimal as GCN accelerators. There are several reasons. 
(i) \emph{GCN applications have highly unbalanced non-zero data distributions:}
non-zeros can be clustered or appear in only a few rows.  This leads to computing challenges 
\cite{gonzalez2012powergraph,abou2006multilevel,latapy2008main} due to workload imbalance \cite{xie2014distributed}. Figure~\ref{power_law} compares the distribution of non-zeros between a typical adjacency matrix in a GCN and a typical sparse-weight matrix in a CNN: the distribution of non-zeros is much more balanced in the SCNN. 
(ii) \emph{Extremely high sparsity}. The sparsity of a graph adjacency matrix often exceeds $99.9\%$, while the sparsity of SCNNs generally ranges from $10\%$ to $50\%$. Therefore, in GCNs the indices of consecutive non-zero elements are often highly scattered; this makes it a major challenge to identify and access sufficient valid non-zero pairs to feed a massive number of PEs per cycle at an acceptable hardware cost (see Section 6). (iii) \emph{Large matrix size.} Real-world graphs are usually very large. For example, the \emph{Reddit} graph has 233K nodes and 115M edges. Its 23K$\times$23K adjacency matrix requires 1.7Tb storage in dense format or 11.0Gb in sparse format, which usually cannot fit into on-chip memory. Although neural networks also have large models, the matrix of a particular layer is often much smaller, e.g., 1k$\times$1k, so the working set often can fit easily into on-chip memory.

\begin{figure}[t] 
\centering
\includegraphics[width=3.5in]{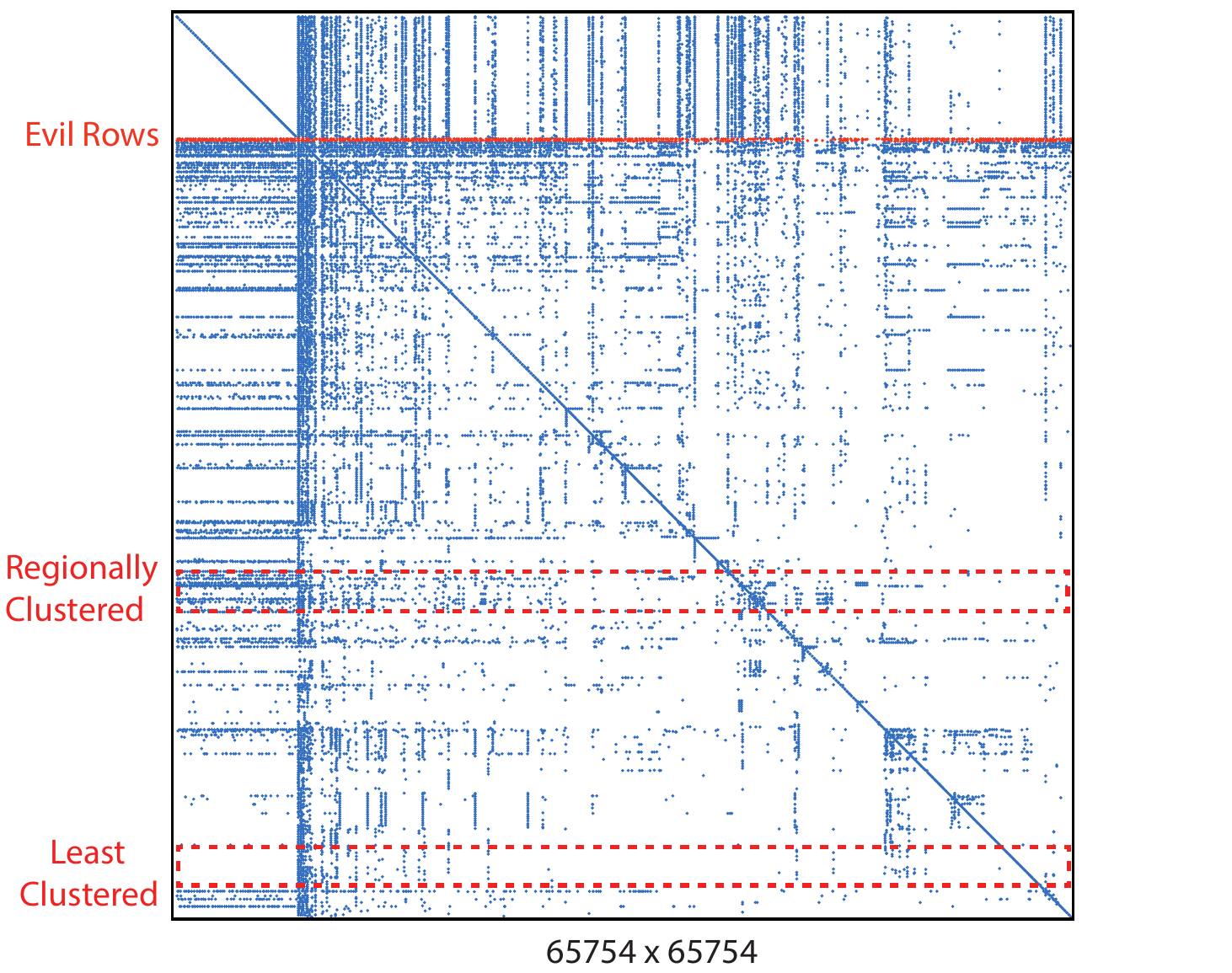}
\caption{Adjacency matrix of NELL following power-law distribution: elements are clustered regionally and in a few rows/cols. The matrix density is 0.0073\%. For better visualization, non-zero dots are enlarged.}
\label{new_distribution}
\end{figure}

For these reasons, novel and efficient accelerator designs are urgently required to accelerate GCN workloads. We therefore propose AWB-GCN, a hardware accelerator for GCN inference with workload auto-tuning. It monitors the workload distribution at three levels at runtime and, accordingly, rebalances the distribution per round\footnote{The calculation of one output matrix column is a round}. 

\begin{figure}[t] 
\centering
\includegraphics[width=3.5in]{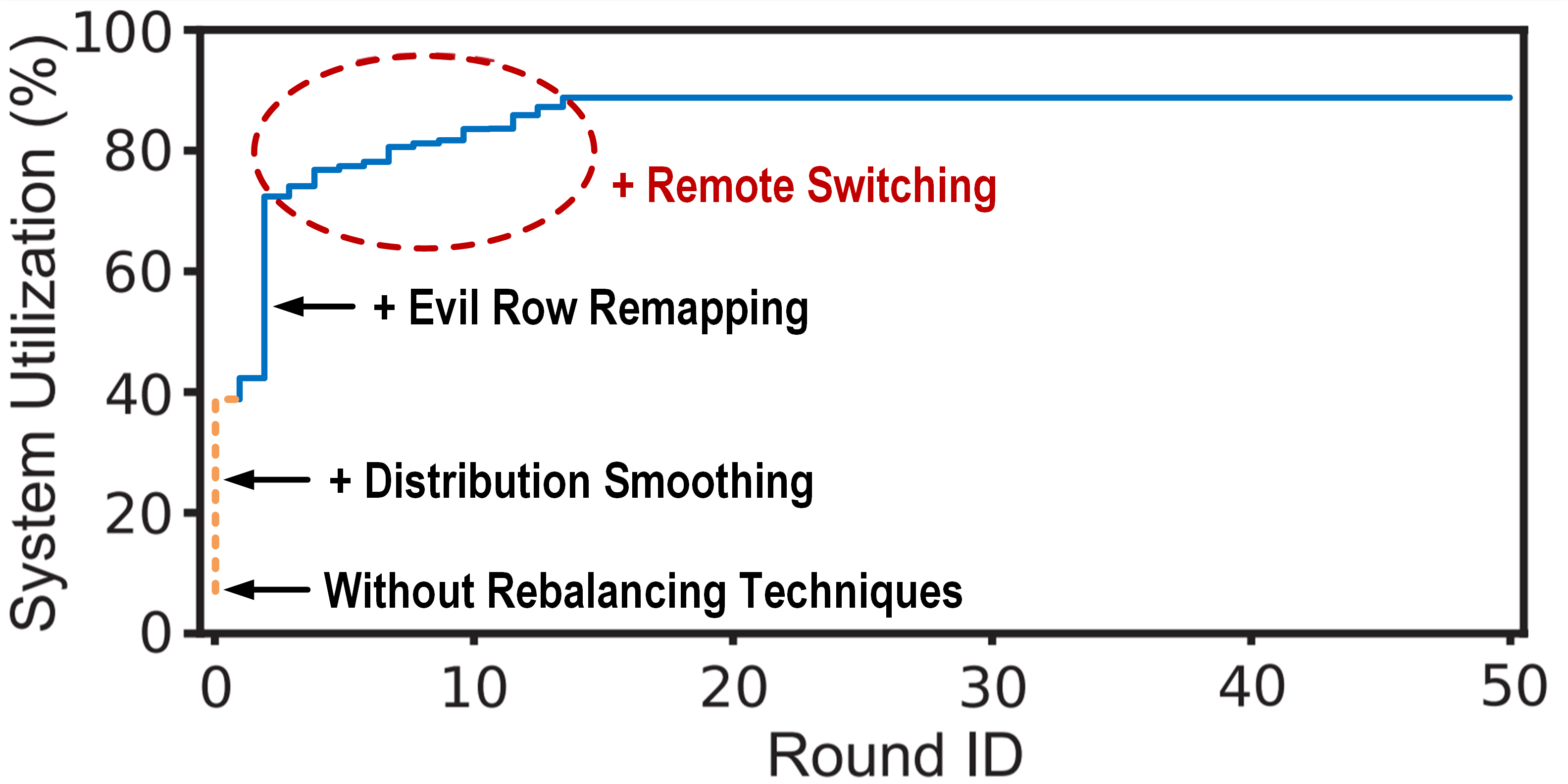}
\caption{AWB-GCN utilization improvement per round.}
\label{autotuning_nell}
\end{figure}

Three techniques are proposed: \emph{distribution smoothing}, \emph{remote switching}, and \emph{evil row remapping}. Distribution smoothing balances the workload among neighbors. In matrices following the power-law distribution, non-zero elements are usually clustered, and, in some cases, appear in just a few rows/columns (Figure~\ref{new_distribution}). Given only distribution smoothing, it would be slow and difficult for an autotuner to converge and achieve good load balance. We solve this problem with \emph{remote switching} and \emph{evil row remapping}. Remote switching shuffles workloads of regions with the most and least clustered non-zero elements, making efficient distribution smoothing possible. If a row is observed to {\it still} contain too many elements to be smoothed or balanced by remote switching, it is designated as an \emph{evil row.} AWB-GCN partitions that row and remaps its non-zero elements to multiple regions (with least clustered elements). Figure~\ref{autotuning_nell} shows the resulting per-round improvement in hardware utilization as these methods are applied (Nell GCN).

This paper makes the following contributions:


\begin{itemize}
\item We propose a novel and efficient architecture for accelerating GCNs and Sparse Matrix Multiplication (SpMM) kernels for matrices with a power-law distribution.
\item To handle the extreme workload imbalance, we propose a hardware-based workload distribution autotuning framework, which includes an efficient online workload profiler and three workload rebalancing techniques.
\item We evaluate AWB-GCN using an Intel D5005 FPGA Acceleration Card with five of the most widely used GCN datasets. Results show that 4K-PE AWB-GCN improves the PE utilization on average by 7.7$\times$ as compared with the baseline without workload rebalancing. Compared with CPUs (Intel Xeon E5-2680v3 + PyTorch Geometric (PyG)), GPUs (NVIDIA Quadro RTX 8000 + PyG), and prior art, AWB-GCN achieves average speedups of 3255$\times$, 80.3$\times$, and 5.1$\times$, respectively.


\end{itemize}


\section{Motivation}



In this section we briefly introduce the GCN algorithm and discuss data characteristics of power-law graphs.


\subsection{Graph Convolutional Network Structure}

Equation~\ref{eq:gcn_layer} shows the layer-wise forward propagation of a multi-layer spectral GCN \cite{wu2020comprehensive, kipf2016semi}: 
\begin{equation}
X^{(l+1)} = \sigma(A X^{(l)}  W^{(l)})
\label{eq:gcn_layer}
\end{equation}
$A$ is the graph adjacency matrix with each row delineating the connection of a vertex with all the other vertices in the graph. $X^{(l)}$ is the matrix of input features in layer-$l$; each column of $X$ represents a feature while each row denotes a node. $W^{l}$ is the weight matrix of layer-$l$. $\sigma(.)$ denotes the non-linear activation function, e.g., \emph{ReLU} \cite{krizhevsky2012imagenet}. In general $A$ needs to be normalized: $\tilde{A} = D^{-\frac{1}{2}} \times (A+I) \times D^{-\frac{1}{2}}$ where $I$ is the identity matrix, and $D_{ii} = \sum A_{ij}$. The reason is that, without normalization, multiplying the feature vector $X^{(l)}$ by $A$ will change its scale: those nodes with more neighbors tend to have larger values under feature extraction. Note that during both training and inference of GCN, $\tilde{A}$ remains constant. Since $\tilde{A}$ can be computed offline from $A$, in the remainder of this paper we use $A$ to denote the normalized $\tilde{A}$. In general $A$ is multiplied only once per layer. However, when multi-hop neighbor information is to be collected, $A$ can be multiplied twice or more (i.e., $A^2$, $A^3$, etc.) per layer.  


Equation~\ref{eq:gcn_layer} is  derived from graph signal processing theory: convolutions on a graph can be converted to a multiplication of signal $x \in R^N$ (i.e., a scalar for each node) and a filter $g \in R^N$ in the frequency domain via the Fourier transform:
\begin{equation}
CONV(g, x) = \mathcal{F}^{-1}(\mathcal{F}(x) \odot \mathcal{F}(w)) = U(U^T x \odot U^T g)
\label{eq:conv}
\end{equation}
where $\odot$ denotes the Hadamard product. $U$ is a collection of eigenvectors for the normalized graph Laplacian $\mathcal{L} = I_N - D^{ - \frac{1}{2}} A D^{ - \frac{1}{2}} = U \Lambda U$. The diagonal matrix $\Lambda$ comprises the eigenvalues. If a frequency domain filter $g_W = diag(W)$ is defined, then Equation~\ref{eq:conv} can be simplified \cite{bruna2013spectral} as:
\begin{equation}
CONV(g_W, x) = U g_W U^T x
\label{eq:conv_simple}
\end{equation}
Equation~\ref{eq:conv_simple} can be further simplified by defining the filter as the Chebyshev polynomials of the diagonal matrix $\Lambda$ \cite{defferrard2016convolutional, kipf2016semi} to obtain Equation~\ref{eq:gcn_layer}.

\begin{figure}[t] 
\centering
\includegraphics[width=3.1in]{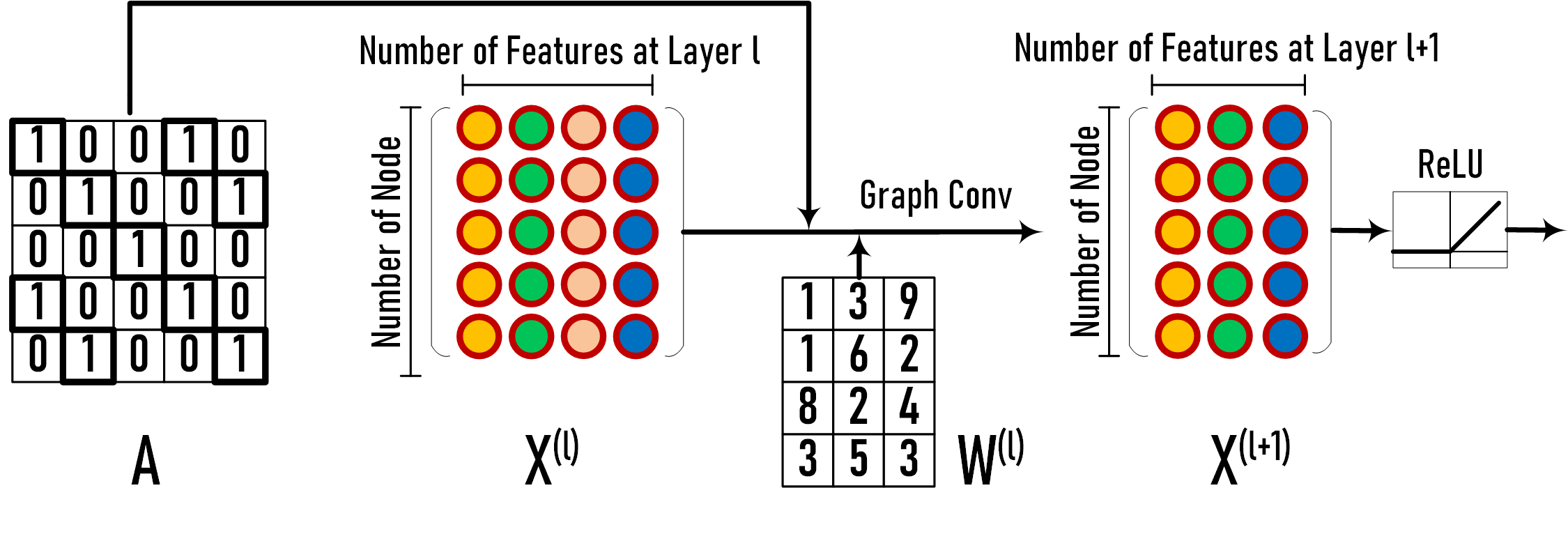}
\caption{Illustration of a GCONV layer in GCNs.}
\label{GCN_illustration1}
\end{figure}

Figure~\ref{GCN_illustration1} illustrates the structure of a Graph Convolutional layer (GCONV). Each GCONV layer encapsulates the hidden features of nodes by aggregating information from neighbors of nodes. By multiplying $A$ and $X^{(l)}$, information from 1-hop connected neighboring nodes are aggregated. By multiplying $AX^{(l)}$ with $W^{(l)}$, and going through the non-linear activation function $\sigma(\cdot)$, we obtain the output of this layer, which is also the feature matrix for the next layer $X^{(l+1)}$. The matrix A will normally be the same in different layers. After multiple layers, the GCN is able to extract very high-level abstracted features for various learning purposes. 




\subsection{Characteristics of Power-Law Graphs}

\begin{figure*}[t] 
\centering
\includegraphics[width=7in]{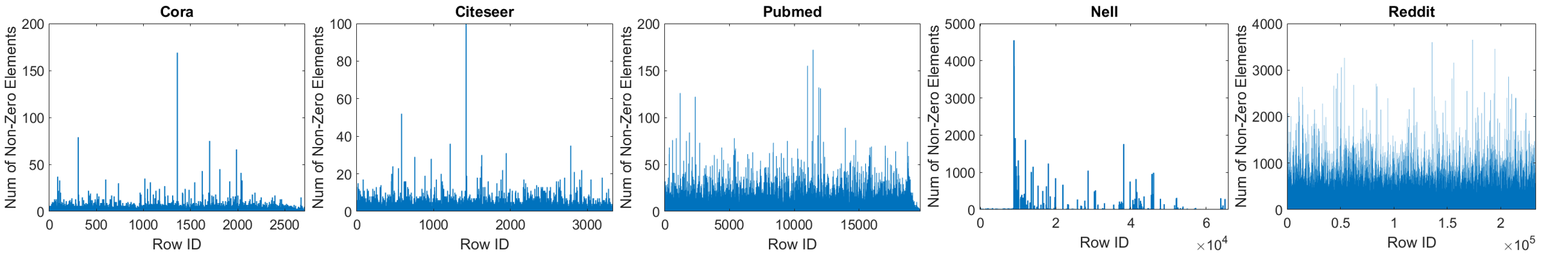}
\caption{Non-zero distribution imbalance of Adjacency matrices in Cora, Citeseer, Pubmed, Nell and Reddit datasets.}
\label{unbalance_cora}
\end{figure*}

Real-world graphs in many critical domains typically follow the \emph{power-law} distribution \cite{xie2014distributed, aiello2001random, chung2004spectra, adamic2001search}, which states that the number of nodes $y$ of a given degree $x$ is proportional to $x^{-\beta}$ for a constant $\beta>0$. This implies that in the adjacency matrix $A$, a small number of rows (or columns) include the majority of non-zeros whereas the majority of the rows (or columns) contain only a few non-zeros but are not empty. Figure~\ref{unbalance_cora} shows the distribution of \emph{non-zero} elements for the five publicly available datasets that are widely used for GCN evaluation \cite{kipf2016semi}. The \emph{power-law} effect is prominent for \texttt{Cora}, \texttt{Citeseer}, \texttt{Pubmed} and \texttt{Nell}. 


Table~\ref{Table:profile_sp_di} lists the density and dimension of matrices in the five GCN datasets used in this paper. Note that adjacency matrix $A$ is always very sparse ($\ge$ 99\%). Matrix $X$ is also usually sparse. For the first layer, the sparsity ($X1$) is usually larger than $90\%$. As the weight matrix $W$ is dense, the output of $AXW$ is also dense. However, because of the $ReLU$ activation function, the final output $X2$ (also the input of the next layer) becomes sparse but with sparsity usually less than $50\%$. The sizes of the matrices in GCNs depend on the dataset and can range from thousands to millions or more. $A$ can be extremely large and is stored in a sparse format. 


\begin{table}[t]
\scriptsize
\centering
\caption{Matrix density and dimensions of 5 widely-used GCN datasets.}
\begin{tabular}{|c|c|c|c|c|c|c|}
\hline
                           &          & CORA  & CITESEER & PUBMED & NELL  & REDDIT \\ \hline
\multirow{4}{*}{Density}  & A        & 0.18\%   & 0.11\%      & 0.028\%   & 0.0073\%   & 0.21\%    \\ \cline{2-7} 
                           & W        & 100\%    & 100\%       & 100\%     & 100\% & 100\%  \\ \cline{2-7} 
                           & X1       & 1.27\%   & 0.85\%      & 10.0\%    & 0.011\%   & 100\%    \\ \cline{2-7} 
                           & X2       & 78.0\%   & 89.1\%      & 77.6\%    & 86.4\%  & 63.9\%   \\ \hline
\multirow{2}{*}{Dimension} & Node     & 2708  & 3327     & 19717   & 65755  & 232965   \\ \cline{2-7} 
                           & Feature & 1433  & 3703     & 500   & 61278  & 602   \\ \hline
\end{tabular}
\label{Table:profile_sp_di}
\end{table}

\section{GCN Baseline Architecture}

This section introduces the multi-core {\it baseline} architecture for GCN acceleration. This baseline supports efficient processing of power-law graphs with ultra-high sparsity and large sizes. This design alone cannot address the workload imbalance issue of power-law graphs, but builds a foundation for its further augmentation, described in the next section, which achieves near-optimal workload balancing.

\subsection{Matrix Computation Order}

To compute $AXW$ at each GCONV layer, there are two alternative computation orders: $(A\times X)\times W$ and $A\times (X\times W)$. The choice is significant as it dictates the volume of non-zero multiplications. Based on profiling, $A$ is ultra sparse and large, $X$ is generally sparse and usually has a large number of columns, and $W$ is small and dense. For $(A\times X)\times W$, since multiplying $A$ and $X$ requires complex sparse-sparse-matrix-multiplication and produces a very large dense matrix, multiplying their product by another dense matrix $W$ leads to significant computation workload and long delay. Alternatively, for $A\times (X\times W)$, both are sparse-dense matrix multiplications (SpMM) and the scale of computation is drastically smaller. Table~\ref{tab:order} lists the amount of computation for the five datasets following the two approaches. Since the difference is quite obvious, in this design we first perform $X\times W$ and then multiply with $A$.




\begin{table}[t]
\scriptsize
\centering
\caption{Operations required under different exec orders.}
\begin{tabular}{|c|c|c|c|c|c|c|}
\hline
                    Layer     &     Order          & CORA     & CITESEER  & PUBMED    & NELL & REDDIT \\ \hline
\multirow{2}{*}{Ops}     & $(AX)W$ & 62.8M & 198.0M & 165.5M &   258G   &   83.3G     \\ \cline{2-7} 
                         & $A(XW)$ & 1.33M  & 2.23M   & 18.6M  &  782M    &  21.4G    \\ \hline
\end{tabular}
\label{tab:order}
\end{table}

\subsection{SpMM Execution Order and Mapping}

We perform column-wise-product-based SpMM \cite{gao2020systematic,deveci2017performance,chen2018performance} as described as follows. Given $S\times B=C$, if $S$ is $(m\times n)$, $B$ is $(n\times k)$, and $C$ is $(m\times k)$, then we can reformulate $C$ as:
\begin{equation}
C=[\sum_{j=0}^{n-1}S_j b_{(j,0)}, \sum_{j=0}^{n-1}S_j b_{(j,1)}, \dots, \sum_{j=0}^{n-1}S_j b_{(j,k-1)}]
\end{equation}
where $S_j$ is the $j$th column of $S$ and $b_{j,k}$ is an element of $B$ at row-$j$ and column-$k$. In other words, by broadcasting the $j$th element from column-$k$ of $B$ to the entire column-$j$ of $S$, we can obtain a partial column of $C$. Essentially, $B$ is processed in a streaming fashion: each element $b_{(j,k)}$ finishes all computation it involves at once and is then evicted. In this way, we reuse the entire sparse matrix $S$ for each column of $C$ ($k$ times in total). To reduce off-chip memory access for matrix $S$, we apply inter-layer data forwarding and matrix blocking techniques (discussed in Section 3.4).

\begin{figure}[t] 
\centering
\includegraphics[width=3.3in]{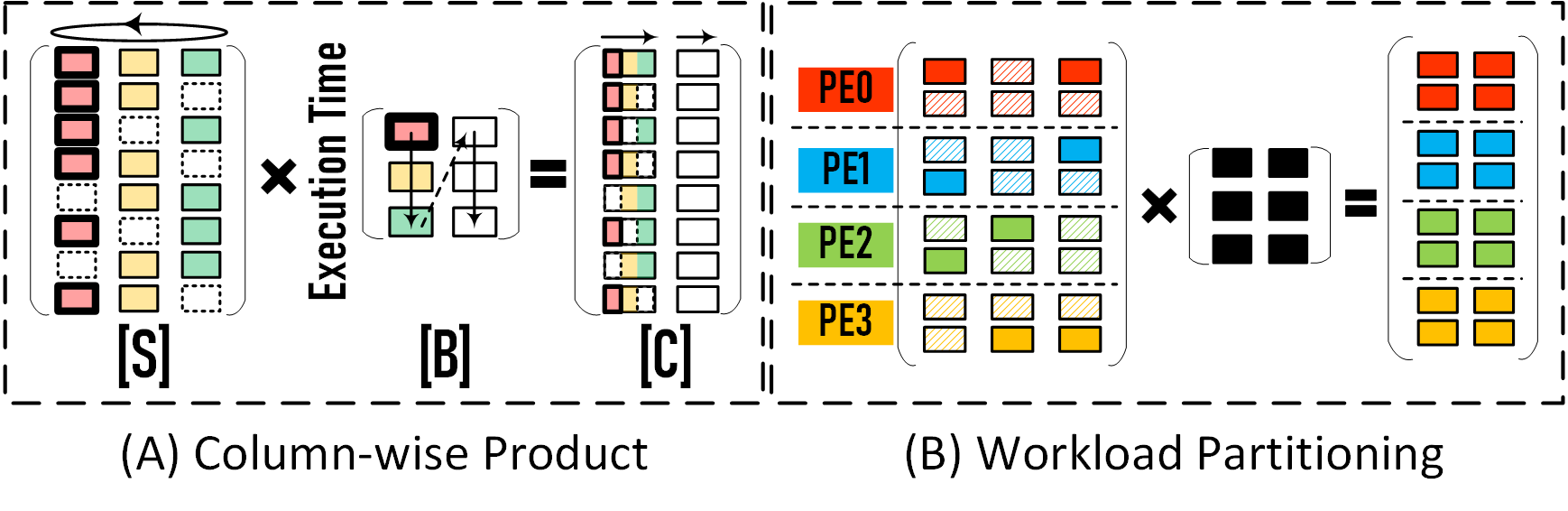}
\caption{(A) SpMM computation order: Column-wise-product; (B) Matrix partitioning \& mapping among PEs.}
\label{SpMM-order}
\end{figure}

This design has additional advantages when $S$ and $C$ are stored in \emph{Compressed-Sparse-Column} (CSC) format. Furthermore, it provides opportunities to pipeline multiple SpMM operations, as is discussed in Section 3.4. Moreover, column-wise-product brings massive opportunities of workload distribution autotuning which is key to achieving high performance. Figure~\ref{SpMM-order}(A) shows the column-wise order for calculating $C$. The columns of $S$ and elements of $B$ in the same color are multiplied and stored as partial results in $C$ with the same color. 

In the baseline design, with the assumption that non-zeros are evenly distributed among the rows, we use a direct and static mapping from matrix rows to PEs to avoid expensive parallel reduction in hardware as illustrated in Figure~\ref{SpMM-order}(B).



\subsection{Design of Baseline Architecture}


Figure~\ref{non-UWB-SpMM-arch} illustrates the baseline design for SpMM calculation with efficient support of skipping zeros. The architecture comprises the modules \emph{sparse-matrix-memory} (SpMMeM), \emph{dense-column-memory} (DCM), \emph{task-distributor \& Queue} (TDQ), \emph{PE-array}, and \emph{accumulation-buffers-array} (ACC Buffer). SpMMeM buffers the input sparse matrix $S$ (from off-chip) and feeds non-zeros and their indices to TDQ. DCM buffers the input dense matrix $B$ and broadcasts its elements to TDQ. TDQ distributes tasks to the PEs. The PE-array performs concurrent multiplication of non-zero pairs, partial result accumulation, and data exchange with the ACC Buffers. Finally, the ACC Buffers cache the partial results of the resulting matrix $C$ for accumulation and send them to the next SpMM engine at the completion of a whole column calculation. Depending on the sparsity and storage format of $S$, i.e., CSC, we have two alternative designs for TDQ:

\begin{figure*}[t] 
\centering
\includegraphics[width=7in]{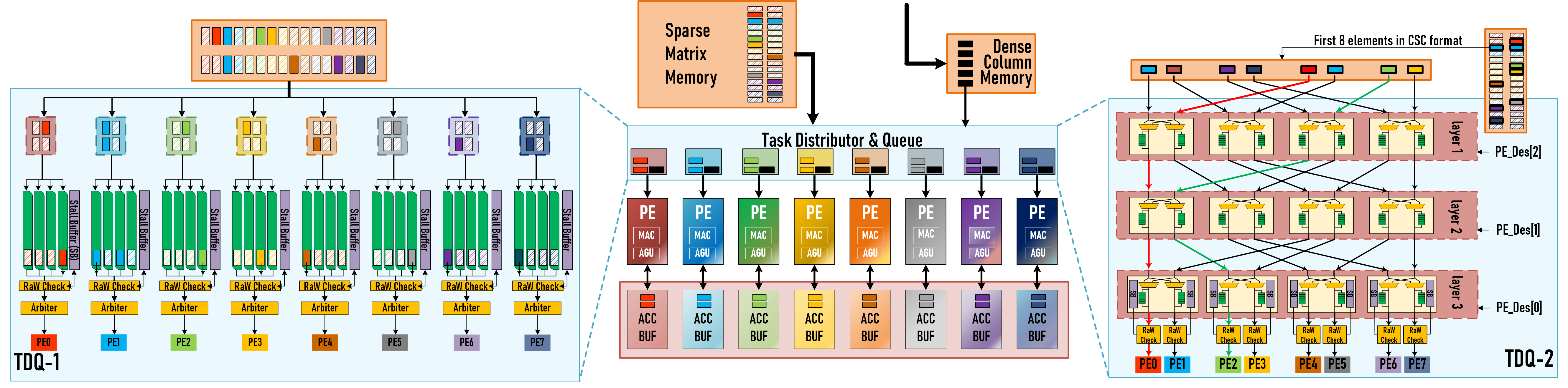}
\caption{Architecture of the proposed baseline SpMM engine.}
\label{non-UWB-SpMM-arch}
\end{figure*}


\vspace{4pt}\noindent\textbf{TDQ-1} (Figure~\ref{non-UWB-SpMM-arch}-left) is used when $S$ is generally sparse (sparsity $< 75\%$) and stored in dense format. We perform the direct row partition as discussed and map non-zeros to the input buffer of the corresponding PEs (Figure~\ref{SpMM-order}(B)). In each cycle, $NPE/(1-Sparsity)$ elements are forwarded to the PE array. Only non-zeros are kept in the queues. Here $NPE$ denotes the number of parallel PEs. Given evenly distributed non-zeros, each PE receives one non-zero per cycle to calculate. In practice, however, the distribution can be very imbalanced and each PE has the chance to receive at most $1/(1-Sparsity)$ in one cycle. Therefore, each PE is equipped with multiple Task Queues (TQs) guaranteeing enough concurrency to cache all valid data. As shown in Figure~\ref{non-UWB-SpMM-arch}-(left), in each cycle a PE can receive up to 4 non-zero elements (sparsity $< 75\%$). Each PE has four task queues to buffer them.

In each cycle, an arbiter selects a non-empty queue, pops an element, checks for a Read-after-Write (RaW) hazard, and forwards it to the PE for processing. Since the computations are all floating-point, the pipelined \emph{multiply-accumulate-unit} (MAC) usually takes several cycles to process, but can still accept new tasks while processing. If the new task tries to accumulate the same partial result of C (i.e., from the same row of A), it actually fetches a stale partial result from the ACC buffer and a RaW hazard occurs. To avoid this hazard, we implement a stall buffer of size T, where T is the delay of the MAC units. We track the row indices currently being processed by the MAC and check whether the current element is targeting the same row in the RaW-check-unit. If so, we buffer that job and delay until the hazard is resolved. The RaW hazard is detected by checking the row index of the coming data.

\vspace{4pt}\noindent\textbf{TDQ-2} (Figure~\ref{non-UWB-SpMM-arch}-right) is used when $S$ is ultra-sparse and stored in CSC format. Since in CSC the non-zeros are contiguous in a dense array, if we can directly process the dense array, we gain from avoiding all the zeros. However, we suffer from the overhead of navigating to the correct PE as the indices of neighboring elements are highly scattered. We use a multi-stage Omega-network for routing the non-zeros to the correct PE according to their row indices. Each router in the \emph{Omega-network} has a local buffer in case the buffer of the next stage is saturated. This design attempts to balance the data forwarding rate and the processing capability of the PEs by sending $NPE$ non-zeros per cycle. This is achieved when non-zero elements are distributed evenly among rows. Compared with a global crossbar network, the Omega-network design scales better and incurs lower hardware complexity. 



When a PE receives a new non-zero pair [$d1,d2$] from TDQ, it (1) performs the new multiplication task with $d1,d2$, (2) fetches the corresponding partial results [$d\_{acc}$] of output matrix $C$ from the ACC buffers according to the newly received row index, (3) accumulates the multiplication result and $d\_{acc}$, and (4) updates the ACC buffers with the new accumulation result. Each PE is coupled with a bank of ACC buffer to store the rows of $C$ it accounts for. A PE has two units: a \emph{MAC} and an \emph{Address-Generation-Unit (AGU)} for result address generation and forwarding. Since $C$ is a dense matrix and stored in dense format, the rows of $C$ are statically partitioned among ACC buffers. Synchronization is only needed when an entire column of the resulting matrix $C$ is completely calculated. 


Overall, for each layer of GCN, we first execute SpMM on $X\times W$. Since $X$ is generally sparse (except the first layer) and stored in dense format, we use TDQ-1. The result of $XW$ is dense. We then compute $A\times (XW)$ which again is SpMM. However, as $A$ is ultra-sparse and stored in CSC format, we use TDQ-2. The result is dense, but after \emph{ReLU}, a large fraction of the entries become zero, and we again have a sparse matrix as the input feature matrix for the next layer. 

\subsection{Pipelining SpMM Chains}

\begin{figure}[t] 
\centering
\includegraphics[width=3.3in]{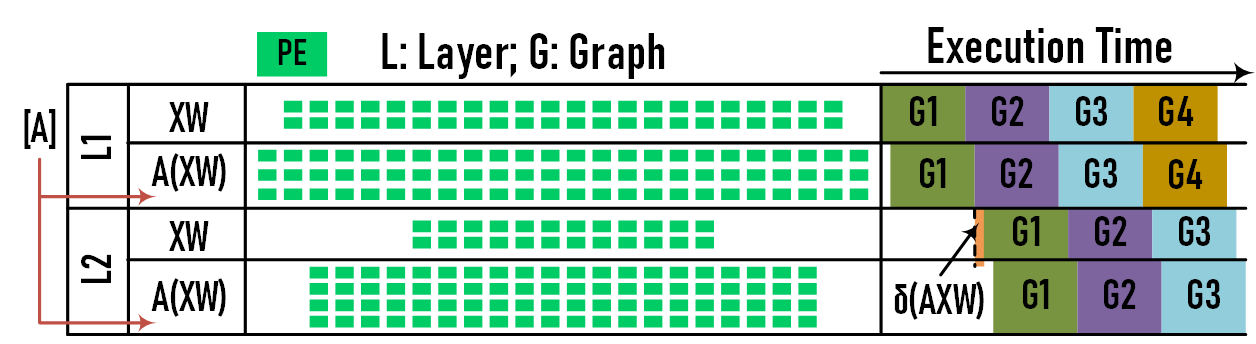}
\caption{Pipelined SpMMs: data production and consumption rates match across consecutive SpMMs by allocating PEs in proportion to workload sizes.}
\label{kernel_parallelism}
\end{figure}

\noindent \textbf{Intra-Layer SpMM Pipelining:} One can exploit the parallelism between consecutive SpMMs (i.e., $X\times W$ and $A\times (XW)$) in a layer through fine-grained pipelining. This is based on the observation that $A$ is constant for the inference of a certain graph. Once a column of $(XW)$ is calculated, we can start the multiplication of this column with $A$ immediately without waiting for the entire $XW$ (see Figure~\ref{kernel_parallelism}). This design has two major benefits: (i) we gain extra parallelism and reduce the overall latency through this fine-grained pipelining, and (ii) instead of requiring off-chip storage to cache the big resulting $XW$ matrix, we only need to buffer a single column of $XW$; this can be done on-chip. This method can be reused within a GCONV layer if $(AXW)$ is left-multiplied by any other sparse matrices. For example, some GCNs collect information from 2-hop neighbors so the layer formulation becomes $A\times(A\times(X\times W))$ and the three multiplications can be pipelined and processed in parallel.

\vspace{3pt}\noindent\textbf{Inter-Layer SpMM Pipelining:} SpMMs from different layers can also be pipelined. To avoid pipeline bubbles and large intermediate buffers, we allocate hardware resources (PEs) in proportion to the workload of each layer (Figure~\ref{kernel_parallelism}). In this way, the output generation of the previous layer matches the data consumption of the current layer, so that the execution time of different layers is similar, given optimal workload balance and PE utilization. Pipelining SpMMs from different layers has two benefits.  First, it exploits inter-layer parallelism. Second, since $A$ is shared for all GCONV layers in the inference of a particular graph, it can be reused by SpMM engines across the layers, so off-chip accesses of $A$ are only required by the first layer. This is done by forwarding elements of $A$ through the layers. 

\begin{figure}[t] 
\centering
\includegraphics[width=2.6in]{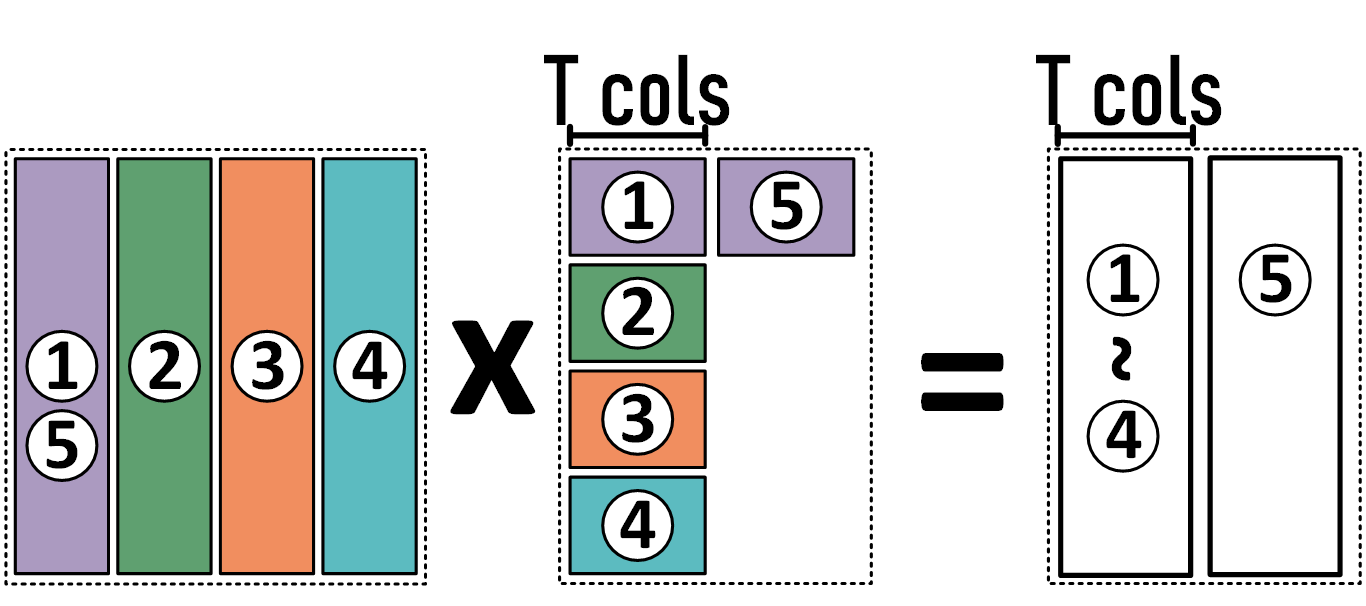}
\caption{Matrix Blocking Optimization to reduce the off-chip bandwidth requirement. The sub-SpMM of each pair of blocks is performed in column-wise-product order. The numbers represent execution orders.}
\label{blocking_matrix}
\end{figure}

\vspace{3pt}\noindent
\textbf{Bandwidth Analysis:} Off-chip data access of the big Adjacency matrix $A$ can be a concern. However, as AWB-GCN always requests and consumes data with continuous addresses, the off-chip memory bandwidth and the burst mode access can be efficiently utilized. Also, we use three extra methods to reduce the off-chip bandwidth requirement: (1) as mentioned above, $A$ is reused across layers; (2) matrix blocking is used to improve the data locality and reuse of matrix $A$. Figure~\ref{blocking_matrix} illustrates how the proposed matrix blocking works without affecting the efficiency of the rebalancing techniques which are discussed in the next section. The numbers in the figure are execution orders. $A$ is partitioned into multiple blocks. Instead of calculating each column of $A(XW)$ by multiplying all blocks of $A$ and the corresponding column of $(XW)$, we calculate $t$ columns of $A(XW)$ in parallel. The calculation of a certain block of $A$ will not start until the previous block is reused $t$ times and finishes calculating its intermediate results of all $t$ columns of the resulting matrix. By doing so, the data reuse of matrix $A$ is improved by $t$ times. Note that this optimization will not hurt the efficiency of the autotuning rebalancing of AWB-GCN, as the sub-SpMM of each block of $A$ is still following column-wise product order. (3) AWB-GCN is equipped with a scratchpad memory to cache parts of $A$ on-chip as much as possible. For example, the $A$ and $X1$ of Cora can be entirely stored on-chip.

Based on our experiments, with the proposed optimizations, the AWB-GCN accelerator requires at most 459 Gbps off-chip bandwidth to keep the hardware busy with 1024 PEs for the 5 datasets evaluated. This bandwidth demand can be generally satisfied by current platforms (e.g., Intel D5005 FPGA board provides 614 Gbps DDR bandwidth; VCU-128 FPGA provides 3680 Gbps HBM bandwidth; NVIDIA V100 provides 7176 Gbps HBM bandwidth).

\subsection{The Workload Balance Problem}

\begin{figure}[t] 
\centering
\includegraphics[width=3.3in]{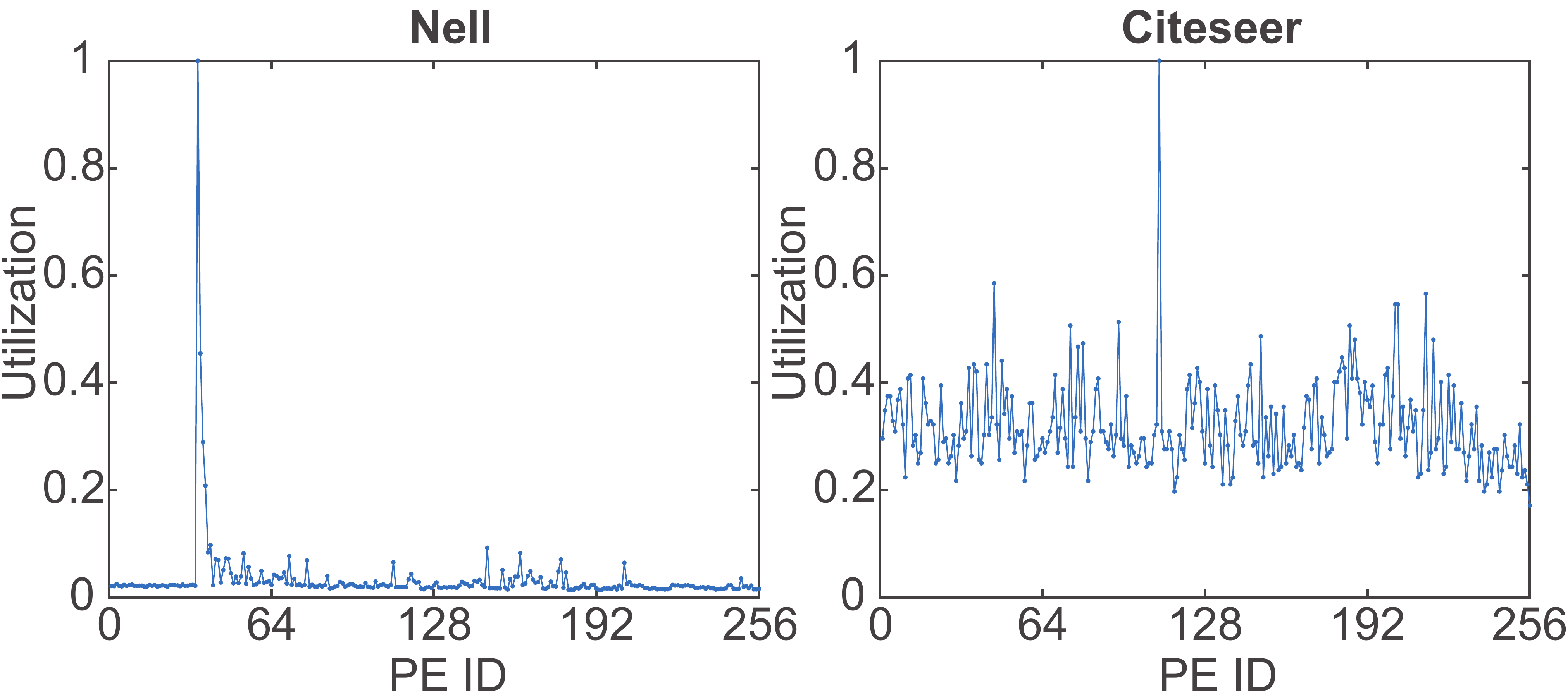}
\caption{PE utilization waves of 256-PE Baseline SpMM engine processing $A\times (XW)$ of Nell and Citeseer.}
\label{imbalanced_utilization}
\end{figure}

The baseline architecture works well when non-zeros are evenly distributed among the rows of $A$. However, when this assumption does not hold, the performance of the baseline architecture can degrade considerably due to workload imbalance among PEs. Figures~\ref{imbalanced_utilization} illustrates the utilization of 256 PEs processing SpMMs with Adjacency matrices of the Citeseer and NELL datasets. As mentioned in Section 1, evil rows and regionally clustered non-zeros in power-law graph matrices bring the inefficiency. The existence of evil rows keeps only a few PEs busy while all others idle most of the time, resulting in significant major crests in the utilization waves; the regionally clustered non-zero elements result in the minor crests; the differences in the numbers of non-zeros in neighboring rows result in other fluctuations. 

A common software approach for dealing with sparse data structures is to profile the structure, e.g., with symbolic analysis, and then use that information to guide the ``real'' processing. For GCNs, however, it has been demonstrated that the preprocessing stage can take 10$\times$ more time than the inference itself \cite{yan2020hygcn}. In this work, we {\it dynamically} adjust hardware configurations for workload rebalancing. This design can be applied to a variety of specialized accelerators for processing sparse data structures.



\section{AWB-GCN Architecture}

In this section, we describe the AWB-GCN architecture. The core is the handling of load balancing at three levels of granularity: \emph{distribution smoothing} for local utilization fluctuations among PEs, \emph{remote switching} for the minor crests, and \emph{row remapping} for the major crests.

\begin{figure}[t] 
\centering
\includegraphics[width=3.5in]{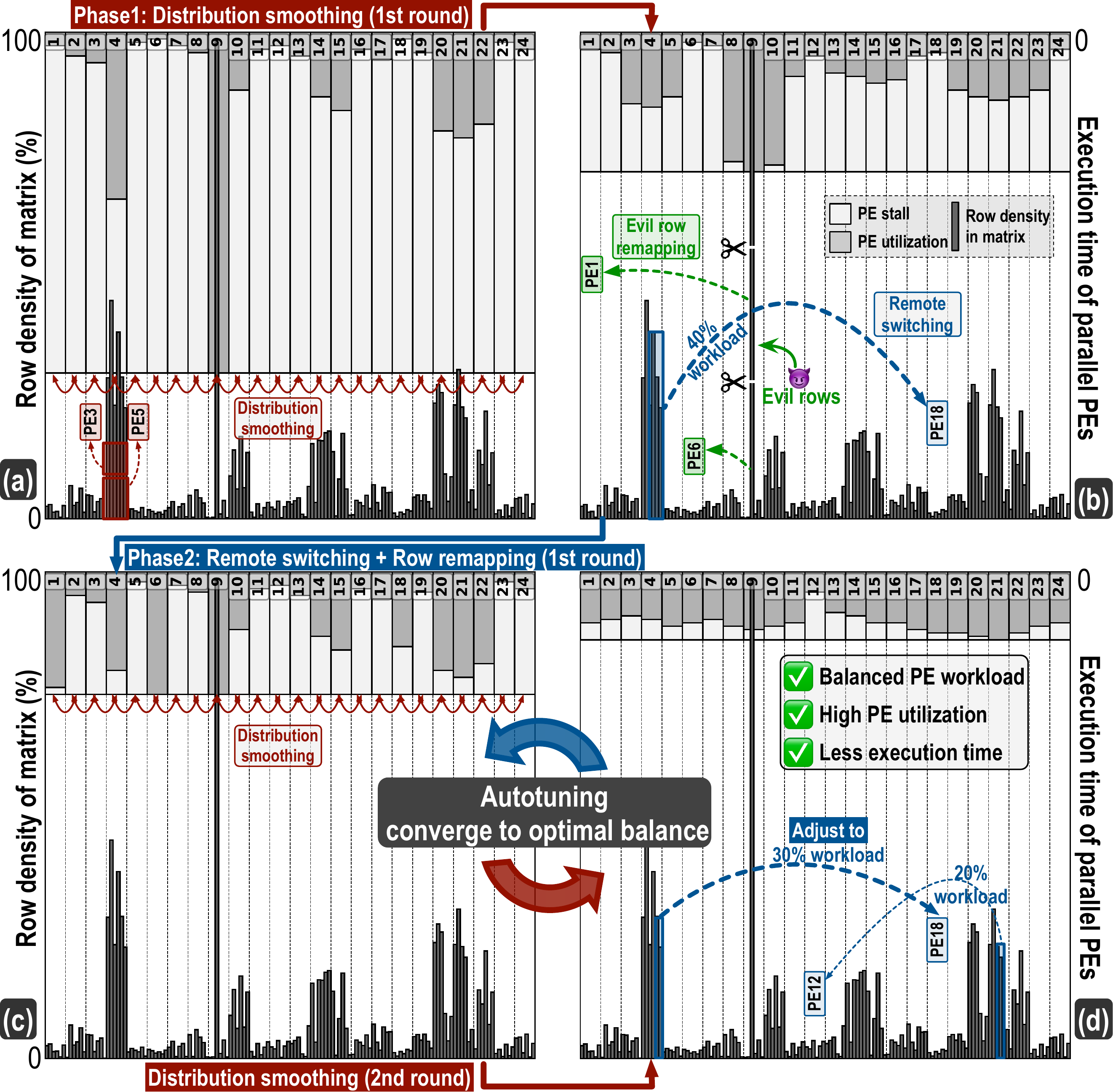}
\caption{Rebalancing process: distribution smoothing, remote switching and row remapping per round. (a)\&(b): 1st round; (c)\&(d): 2nd round.}
\label{uw_example}
\end{figure}

Figure~\ref{uw_example} illustrates autotuning with 24 PEs performing SpMM on a power-law matrix. The gray bars at the top show the execution time of parallel PEs; the length changes dynamically through the process. The narrower bars at the bottom show the static density-per-row of the matrix. Ideally, at the end of autotuning, all bars on the top becomes short and have the same length. Each round of autotuning includes two phases: First, data processing and distribution smoothing in phase 1; then remote switching and row remapping.

Figure~\ref{uw_example} (a)\&(b) illustrate the first round of autotuning. The progression from Figure (a) to (b) shows the first phase. Figure (a) gives estimated execution time without distribution smoothing; Figure (b) shows the actual execution time with distribution smoothing applied. During phase 1, PEs keep offloading workloads to their less busier neighbors, resulting in a more flat and smooth execution time wave (shown in (b)). Meanwhile, the execution time of PEs at the wave crests and troughs is recorded by the Autotuner. 

After all the PEs have finished, phase 2 starts. The Autotuner partitions and remaps evil rows to PEs at troughs and switches workloads of the PEs at the minor crests with the ones at the troughs. The green and blue arrows in (b) show evil row remapping and remote switching decisions, respectively. After these decisions are made, the second round of autotuning starts (Figures (c)\&(d)). With remote switching and row remapping determined in the first round, the initial workload distribution among PEs at the start of the second round (shown in (c)) can be more efficiently balanced by distribution smoothing (shown in (d)). The blue arrows in (d) show that remote balancing not only finds new pairs of PEs to switch workloads, but also adjusts the switch fractions determined in the previous round. After several rounds, the system converges to optimal balanced status; this is then used for the remainder of the computation.

All profiling and adjustment are performed at runtime. We now present design details.

\subsection{Distribution Smoothing}


At the start of processing, rows are evenly distributed among PEs as introduced in Section 3.2 (as shown in Figure~\ref{SpMM-order}(B)). During the calculation of each round, we employ distribution smoothing by averaging out the workloads among neighbors. The architecture is able to monitor the runtime PE utilization information by tracking the number of pending tasks in TQs and keep offloading the work of PEs with more pending tasks to their less busy neighbors. However, the offloaded work needs to be returned for aggregation after processing. Due to chip area and design complexity restrictions, we may offload workloads among direct neighbors, 2-hop neighbors, or even 3-hop neighbors, but not farther ones. 

\begin{figure}[t] 
\centering
\vspace{-0pt}
\includegraphics[width=3.5in]{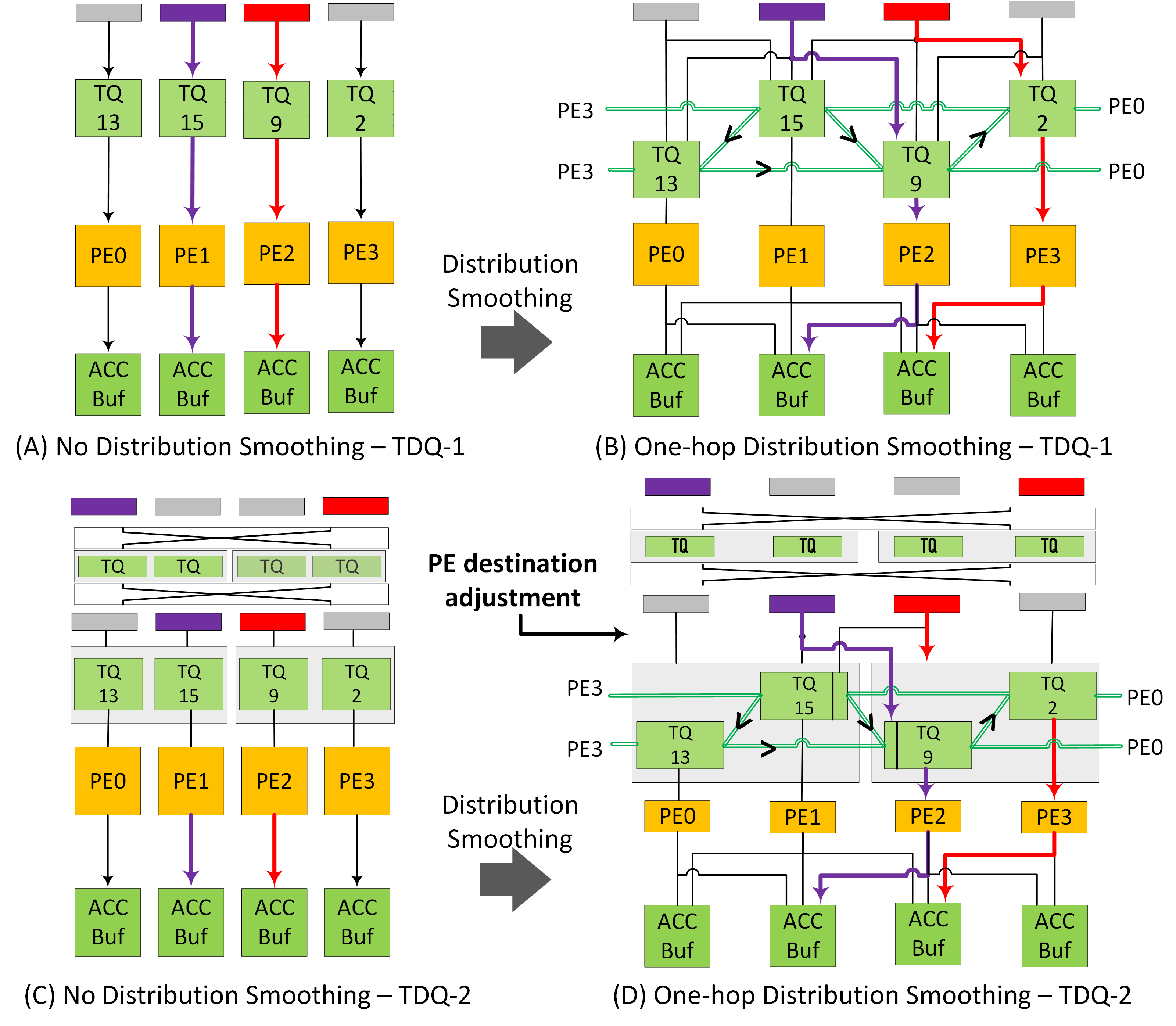}
\caption{Simplified architecture of distribution smoothing.}
\label{uw1_arch}
\end{figure}

Figure~\ref{uw1_arch} illustrates the hardware design of 1-hop distribution smoothing for TDQ-1 and TDQ-2.

\vspace{3pt}\noindent\textbf{TDQ-1:} Before a new task is pushed into the TQ of a PE, the PE compares the number of pending tasks with those in the neighboring TQs. The task is then forwarded to the TQ with the fewest pending tasks. If forwarded to a neighbor, the result needs to be returned to the ACC buffer of its original PE after accumulation (see Figure~\ref{uw1_arch}-(B)). The calculation of valid return address and accumulation of partial results are done in the neighbor PE.

\vspace{3pt}\noindent\textbf{TDQ-2:} The final layer of the multi-stage Omega network handles neighbor task forwarding. As shown in Figure~\ref{uw1_arch}-(C) (also in Figure~\ref{remote_arch}), multiple PEs share the same final-layer switch; we refer to these PEs as a {\it group}. AWB-GCN keeps tracking the usage of TQs of the final layer. Once a new task is forwarded to the final-layer switch, the TQ usages among neighbors are compared and then the task is routed to the PE with the lowest TQ usage. To enable PEs on the group edge (i.e., the leftmost or rightmost PEs per group) to communicate with their out-of-group neighbors, we augment the Omega-network by adding 2 extra links per switch in the final layer, as shown in Figure~\ref{uw1_arch}-(D). Note that Figure~\ref{uw1_arch}-(D) shows sharing only among 1-hop neighbors. By considering more distant hop neighbors, a more balanced design is obtained at the cost of higher hardware complexity and area. This is discussed in the evaluation section.

\vspace{3pt} Distribution smoothing helps remove local utilization fluctuations (Figures~\ref{uw_example}(a) to (b)), but is not sufficient when (1) non-zeros are clustered in a region across many PEs, so that neighbors are mostly busy and have no chance to help each other, resulting in a minor utilization crests (PE20,21,22 in Figure~\ref{uw_example}(b)); or (2) most non-zeros are clustered in only a few rows so that the major crests cannot be eliminated even if all neighboring PEs help (PE9 in Figure~\ref{uw_example}(b)).

\subsection{Remote Switching}

To address regional clustering, we propose remote switching. This process partially or completely exchanges the workloads between under- and overloaded PEs, i.e., at centers of utilization wave troughs and crests, respectively. The switch fraction is determined at runtime by an autotuner and is based on per-round PE utilization. 
As the sparse matrix $A$ is reused during the processing per round, the switch strategy generated in prior rounds is valuable in the processing of later rounds. The accelerator remembers the switch strategies used in the current round and incrementally optimizes them based on the utilization information obtained in the next round. In this way, remote switching is able to flatten the crests and troughs; after several rounds of autotuning, the switch strategy best matching the sparse structure of $A$ is obtained, and is used for the remaining rounds for almost perfect PE utilization.



\begin{figure*}[t] 
\centering
\includegraphics[width=7.2in]{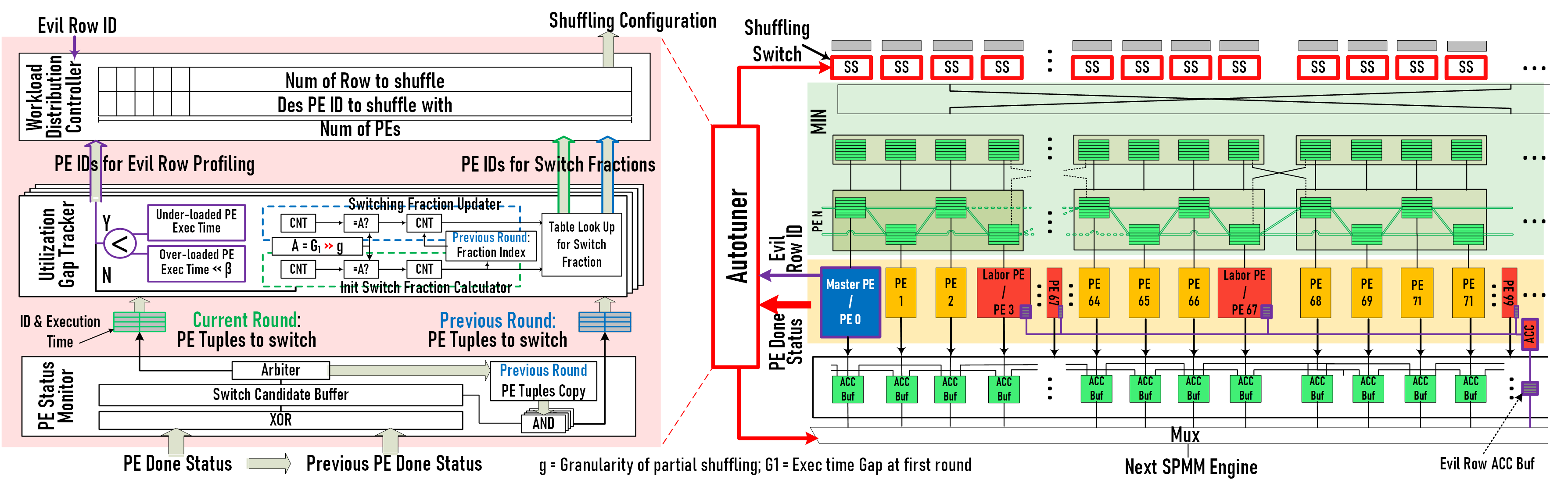}
\caption{Overall architecture of SpMM engine in AWB-GCN with three rebalancing techniques: distribution smoothing, remote switching (red bordered) and evil row remapping (purple bordered). Here every 128 PEs has one Super-PEs and four Labor-PEs. These numbers can be customized.}
\label{remote_arch}
\end{figure*}

The hardware design is shown in Figure~\ref{remote_arch}. The over-loaded and under-loaded PEs are identified by using the \emph{PE Status Monitor} (PESM) of Autotuner during Phase one of autotuning. Recall that each TQ has a counter to track the number of pending tasks; these can trigger an {\it empty} signal when reaching zero. These empty signals are connected to the PESM. At each cycle, the updated empty signals are $xor$ed with their values recorded on the previous cycle. The XOR results indicate which PEs are newly finished; this information is stored in the \emph{Switch Candidate Buffer}. 

At the start of each round (after $A$ is totally sent to TDQs), the arbiter scans the buffer and record IDs of newly done PEs until enough under-loaded PEs have been found. The number of PE tuples for switching at each round can be customized. In Figure~\ref{remote_arch}, four tuples of the most over- and under-loaded PEs are selected for remote switching. After the arbiter finds the first 4 idle PEs, it stops scanning the buffer and instead waits for the completion signal ($bit-AND$ all empty signals) from the system, which implies all PEs have become idle. Meanwhile, the Switch Candidate Buffer caches the newly idle info of the most recent cycles. Whenever the arbiter receives the completion signal it starts to scan the buffer and continues until the four most over-loaded PEs have been found. Note that the arbiter does not select neighbor PEs continuously; this guarantees that PE tuples selected by PESM are at different crests and troughs of the utilization wave.

To avoid thrashing, we only exchange a portion of the workload between PEs. We use the following equation to calculate the number of jobs (i.e., rows of $A$) to be switched in the $i$-th round (i.e., a column of $B$), $N_i\_init$:
\begin{equation}
N_i\_init = G_i/G_1 \times (R/2)
\label{eq:balance}
\end{equation}
where $G_i$ is the workload gap of the selected PE tuple at the $i$-th round, and $R$ is the number of rows per PE under equal mapping.  Here, workload gap is approximated as the difference of execution cycles to finish all tasks.

In the $i+1$-th round, new PE-tuples are selected and their switch fractions are calculated. Meanwhile, the autotuner also tracks the post-switching utilization gaps of PE-tuples selected in the prior rounds and uses them as feedback to adjust the switch fraction $N_i\_init$; this minimizes the utilization gaps further. The workload switching fraction for each tracked PE-tuple is adjusted for two or more rounds and is highly likely to converge to the optimal distribution. 
Equation~\ref{eq:balance} can now be rewritten as follows:
\begin{equation}
N_{i,j} = 
\begin{cases}
G_i/G_1 \times (R/2) \quad &if\ j=0\\
N_{(i-1),(j-1)} + G_i/G_1 \times (R/2)\quad &if\ j>0
\end{cases}
\label{eq:balance1}
\end{equation}
where $j$ denotes the number of rounds of fraction update. $N_{i,j}$ indicates that the current PE-tuple is in its $j$-th update and its initial fraction to switch was calculated in the $i-j$-th round. The number of rounds tracked simultaneously can be customized and depends on the size of the tracking window in the PESM; this is an area/performance tradeoff. In Figure~\ref{remote_arch}, two consecutive rounds are tracked.

Calculation of Equation~\ref{eq:balance1} is done in the \emph{Utilization Gap Tracker} (UGT in Figure~\ref{remote_arch}). To reduce the hardware cost of calculating $G_i/G_1\times (R/2)$, we use a hardware-friendly approximation with threshold-based counting and table lookup; when the most under-loaded PE is found, the left CNTs in UGT start counting. The execution cycle gap (G1) at the first round is right-shifted by $g$ bits (the granularity for division approximation). The result is used as a threshold. Whenever the left CNTs reach the threshold, they get back to 0 and the right CNT adds 1. When the most over-loaded PE is found, the counting stops. Assuming the right CNT counts to $q$, we know the execution time gap at the current round is approximately $q\times G1/(2^{g})$.

Using $q$ as the address to access the Table for Switch Fraction Lookup (TfSFL), we know the approximate number of rows that needs to be switched. More details are omitted due to space limitations. Once the number of rows to be switched is known, it is forwarded to the \emph{Workload Distribution Controller} (WDC) together with the corresponding PE IDs. At the start of the next round, the destination PE of these rows is updated in the \emph{Shuffle Switches} (SS). By doing so, the non-zeros in these rows will be forwarded to the post-switching PEs in the coming rounds.

Furthermore, in order to reduce the workloads of overloaded PEs more efficiently, all operations related to the main-diagonal elements at the rows assigned to these PEs are skipped during processing. Instead of performing these operations, when the required elements of the dense matrix reach TDQs, they will be directly forwarded to the ACC Buffers of the post-switching PEs and be accumulated just before the final accumulation results are sent to the next kernel.

Remote switching followed by distribution smoothing is efficient on getting rids of most of crests of utilization waves. However, for the major crests resulted from evil rows which have too many non-zeros to be shared only by neighbors, extra effort is required.

\begin{figure*}[!htbp] 
\centering
\includegraphics[width=7in]{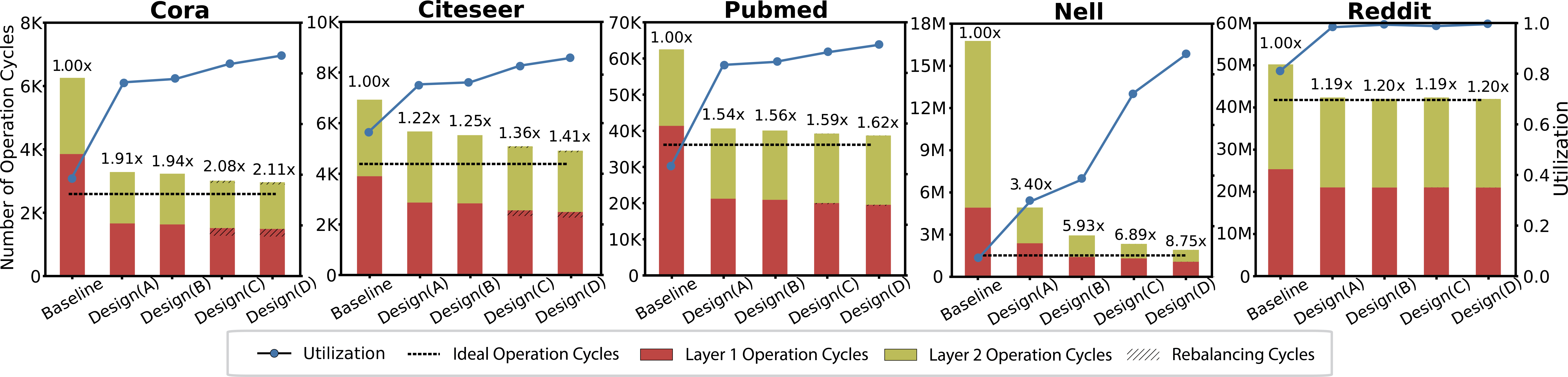}
\caption{Overall performance and PE utilization of 1K-PE AWB-GCN with five design choices.}
\label{eva(A-E)}
\end{figure*}

\begin{figure*}[!htbp] 
\centering
\includegraphics[width=7in]{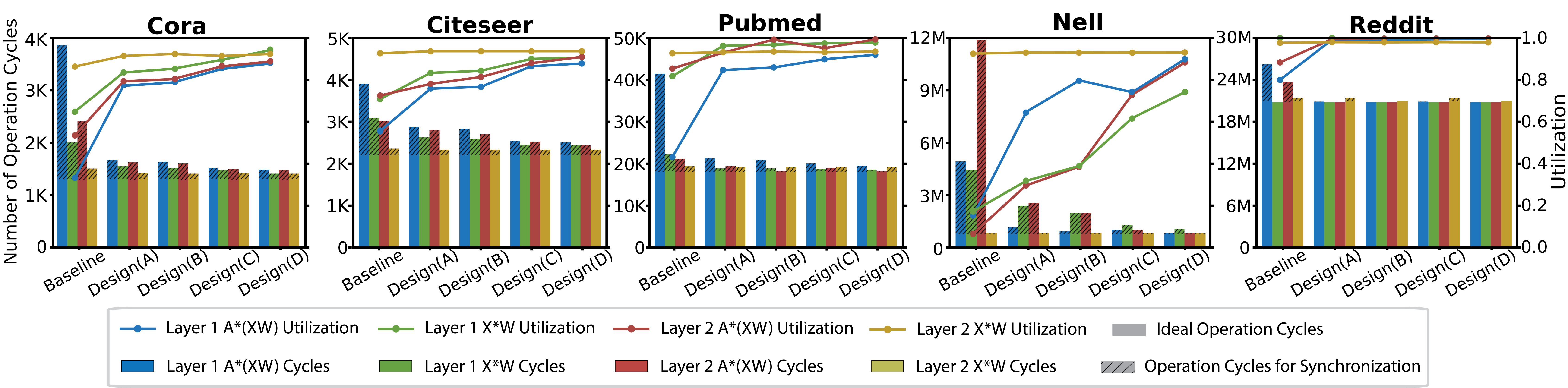}
\caption{Per-SpMM performance and PE utilization of 1K-PE AWB-GCN with five design choices.}
\label{eva(F-J)}
\end{figure*}

\subsection{Evil Row Remapping}

We address evil-row clustering by building row remapping support into the remote switching hardware. With row remapping, the evil row is distributed to the most under-loaded PEs in troughs; in this way the neighbors of these PEs can help. Row remapping is triggered based on demand at the end of each round. The autotuner calculates the utilization gaps between the most over- and under-loaded PEs and determines whether their gaps are too big for remote switching to handle. If yes, row remapping is performed. The workloads of the PE overloaded in the current round are switched (temporarily) with a \emph{Super-PE} in the next round. During processing of the next round, the Super-PE counts the numbers of non-zeros per row and finds the evil rows containing the most non-zeros. In the round after, the workloads of each evil row are partitioned and distributed to a set of \emph{Labor-PEs} controlled by the Super-PE.


After evil rows are remapped to labor-PEs, the original workloads of the labor-PEs can still be swapped with the most under-loaded PEs via remote switching; this ensures that even if the labor-PEs are overloaded originally, they do not become new crests after row remapping. If a labor-PE itself is found to have an evil row, evil row remapping will first map its workload to the master-PE, and then distributively remap the evil row back to labor-PEs, including the one which has the evil row originally. By remapping evil rows statically to certain PEs instead of dynamically to random ones, the aggregation of partial results becomes hardware efficient. If row remapping is not triggered, Super- and Labor-PEs serve as regular PEs.

The existence of evil rows is generally the most critical bottleneck, especially when utilization is lower than 50\%. The proposed row remapping technique makes it possible for the autotuner to find the optimal workload distributions and achieve high utilization. As evil row remapping is normally triggered during the first few rounds, the utilization of the system increases rapidly right at the start and the autotuner generally converges quickly.

Figure~\ref{remote_arch} illustrates the hardware support of row remapping. For clarity, only one Super-PE and its four Labor-PEs are shown. The Labor-PE has an architecture similar to the normal PE, but they are connected to an adder tree for result aggregation. The aggregated results of evil rows are cached in a small separate ACC buffer. The super-PE is much bigger than other PEs, as it serves as a profiler to find the evil rows. It is equipped with two extra modules: a parallel sorting circuit that tracks the rows with the most non-zeros; and a non-zero counter (including a local buffer) that records the number of non-zeros per row. Workload remapping between Super-PE \& Labor-PEs and workload switching between Super-PE \& the PE-with-evil-rows are handled by augmenting the Autotuner as follows. First, the UGT module is equipped with a comparator to identify whether evil row remapping is required; if it does, then the UGT will send the information to WDC. The WDC knows the IDs of the Super-PE and Labor-PEs. If row remapping is triggered or an evil row is found, the entries of the Super- and Labor-PE at Distribution Switch Table in the WDC are updated. This enables workload switching and remapping in the coming round.

\section{Evaluation}

In this section, we evaluate AWB-GCNs with different design choices and compare them with other platforms processing the same networks.

\subsection{Evaluation Configuration}

We implement AWB-GCNs in Verilog HDL and measure PE utilization, performance, energy efficiency, and hardware resource consumption on Intel acceleration card D5005 which is equipped with a Stratix 10 SX FPGA. Note that the FPGA is only used as an evaluation platform to demonstrate the performance of AWB-GCN. The design is a general architecture that does not leverage any FPGA-specific features.

To measure utilization, we add a counter to each PE to track the number of idle cycles. The number of operating cycles (i.e., execution delay) is also measured with a hardware counter. The hardware consumption and operating frequency are reported by Quartus Pro 19.4 after synthesis and implementation. To perform fair cross-platform comparisons, we implement GCNs with the state-of-the-art and famous software framework, PyG\cite{fey2019fast}, and run them on Intel Xeon E5-2680-V3 CPU and NVIDIA RTX 8000 GPU. We also compare AWB-GCN with prior work on GCNs such as HyGCN\cite{yan2020hygcn}.

The datasets used for evaluation are \emph{Cora}, \emph{Citeseer}, \emph{Pubmed}, \emph{Nell} and \emph{Reddit}; these are the five most widely used publicly available datasets in GCN research.




\subsection{AWB-GCN Evaluation}



\begin{figure}[t] 
\centering
\includegraphics[width=3.5in]{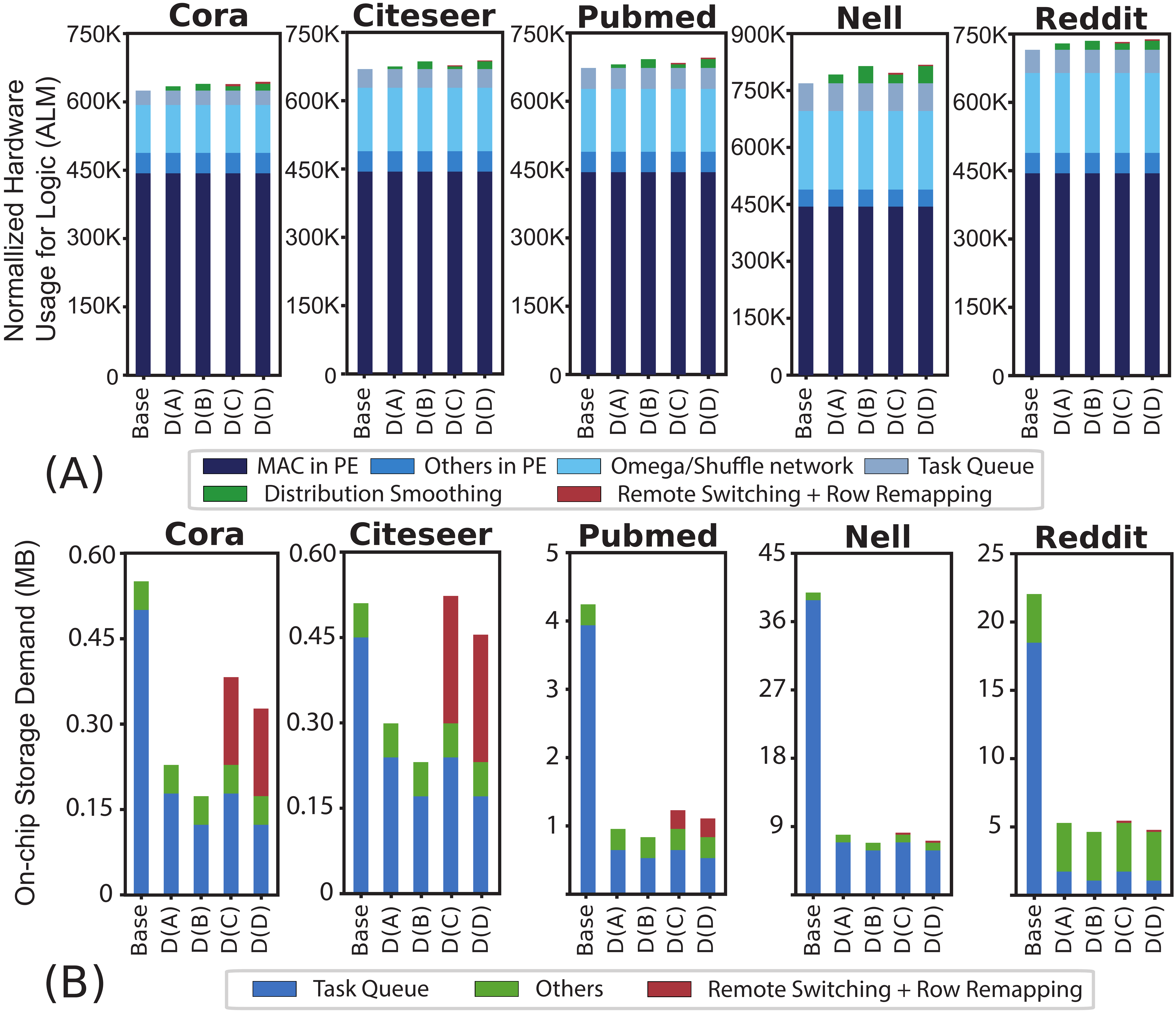}
\caption{(A) Hardware resource consumption normalized to the number of ALMs and (B) On-chip storage demand of 1K-PE AWB-GCN.}
\label{eva(K-O)}
\end{figure}

\begin{figure*}[!htbp] 
\centering
\includegraphics[width=7in]{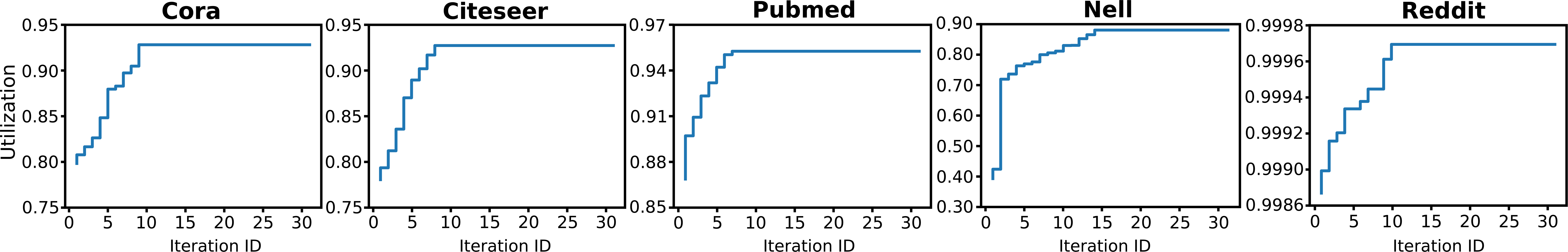}
\caption{AWB-GCN PE (1K) average utilization per round of workload autotuning.}
\label{autotuning}
\end{figure*}

\begin{figure*}[!htbp] 
\centering
\includegraphics[width=7in]{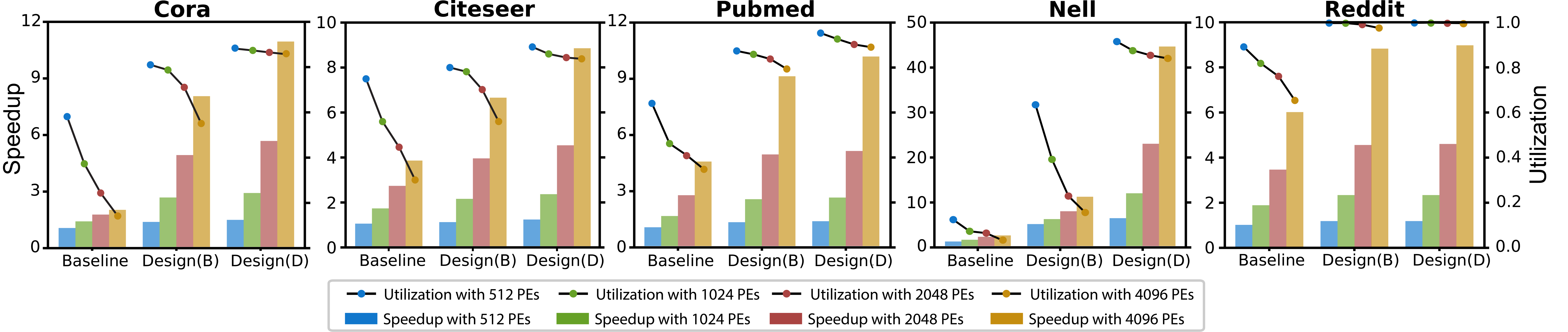}
\caption{Scalability evaluation: PE utilization and overall performance of Baseline, Design(B) and Design(D) of AWB-GCNs with 512, 1K, 2K and 4K PEs.}
\label{eva2}
\end{figure*}

Design efficiency is evaluated by comparing the performance, hardware resource consumption, and PE utilization of the 1K-PE baseline design without any rebalancing techniques (i.e., \emph{\textbf{Baseline}}) with the four different design choices of 1K-PE AWB-GCNs: (i) 1-hop distribution smoothing (i.e., \emph{\textbf{Design(A)}}), (ii) 2-hop distribution smoothing (i.e., \emph{\textbf{Design(B)}}), (iii) 1-hop distribution smoothing plus remote switching and row remapping (i.e., \emph{\textbf{Design(C)}}), and (iv) 2-hop distribution smoothing plus remote switching and row remapping (i.e., \emph{\textbf{Design(D)}}). The only exception is for \emph{Nell} where we use 2-hop and 3-hop distribution smoothing (rather than 1-hop and 2-hop) due to its extremely clustered distribution.


Figure~\ref{eva(A-E)} compares the end-to-end GCN inference latency and average utilization of PEs for the five designs over the five datasets. The lines show the overall PE utilization. The bars show the breakdown of execution cycles of different GCN layers. The latency of ReLU is too low to show in the figure. The off-chip memory access latency is overlapped with computation. We also mark the latency lower bound assuming theoretically ideal PE utilization. For \emph{Cora}, \emph{Citeseer}, \emph{Pubmed}, \emph{Nell} and \emph{Reddit}, comparing to Baseline, Design(B) can improve PE utilization from 38\%, 56\%, 44\%, 7.1\% and 82\%, to 79\%, 77\%, 86\%, 39\%, and 99\%, respectively, leading to $1.94\times$, $1.25\times$, $1.56\times$, $5.93\times$, and $1.19\times$ performance improvement. Enabling remote switching can further improve PE utilization to 88\%, 88\%, 93\%, 88\%, and 99\%, bringing performance gain to $2.11\times$, $1.41\times$, $1.62\times$, $8.75\times$, and $1.20\times$. The results show that AWB-GCN always provides high utilization and close to theoretical peak performance for datasets with various levels of power-law distribution.

In AWB-GCN, hardware resources allocated to different layers are in proportion to their volume of operations. Thus, when perfect utilization is achieved, the same execution delay is observed for all layers. As shown in Figure~\ref{eva(A-E)}, the green and red bars have similar lengths at Design(D), while their lengths vary significantly for the Baseline. 

The shaded area in Figure~\ref{eva(A-E)} represents the performance overhead of the proposed rebalancing techniques. Distribution smoothing is performed during the processing of PEs incurring no overhead so Designs(A)\&(B) are not shaded. For Designs(C)\&(D), most of the tasks for remote switching and row remapping are also performed in parallel with the processing of PEs, e.g., all tasks at PESM and the utilization gap calculation at UGT. However, the table lookup for switch fraction at UGT and data update at WDC must be done sequentially between the processing of two consecutive iterations (columns). They introduce negligible overheads (shaded area of bars for Design(C)\&(D)), before the system converges to optimal balanced status. The shaded areas are only visible for Cora and Citeseer whose workloads are relatively lighter.

Figure~\ref{eva(F-J)} further breaks down the numbers of execution cycles and shows results for every SpMM kernel; this demonstrates the benefits of AWB-GCN on kernels with various sparsity, size and distributions. The shaded area of the bars represents the {\it Sync} cycles due to workload imbalance; the unshaded area represents the {\it Ideal} cycles assuming perfect workload balance. The bars in different colors represent the execution cycles of the four SpMM kernels in the two-layer GCNs \cite{kipf2016semi, wu2020comprehensive}: \emph{$A\times(XW)$ and $X\times W$ at Layer 1 and 2}. The lines show the corresponding PE utilizations. As shown in Figure~\ref{eva(F-J)}, Design(D) significantly minimizes the synchronization overheads for all kernels of the 5 GCN models. 



Comparing SpMM kernels, utilization improves significantly for $A\times(XW)$ at both layers and $X\times W$ at Layer-1. As for $X\times W$ at Layer-2, although $X$ is also sparse after activation is performed, its sparsity is much lower than that of the $X$ at Layer-1 and its non-zero distribution does not follow the power-law (similar to that of the sparse matrices in SCNNs); utilization is thus high even with the baseline design.

Figure~\ref{eva(K-O)}(A) compares the hardware resource usage of the five designs over the five datasets. To show comparable breakdowns to ASIC implementations, the results of hardware resource usage are reported by Quartus Pro 19.4 after synthesis (therefore, FPGA-specific optimizations are not included) and are normalized to the number of \emph{Adaptive Logic Modules} (ALMs). ALM is the basic component of Intel FPGAs. In an ASIC design the analogue would be the number of transistors. The blue segments represent the resource usage for the modules of the baseline design including MAC units in PEs, other modules in PEs, task queue control logic, and omega/shuffle networks. The green and red segments refer to the hardware overheads for the support of \textit{distribution smoothing} and \textit{remote switching + row remapping}. Note that in practical FPGA implementations, MAC units in PEs are instantiated with floating-point DSP slices. In order to show the area breakdown more clearly, we normalize the DSP slices to ALMs. As shown in Figure (A), the overheads of 1-hop and 2-hop distribution smoothing are on average 3.5\% and 6.7\%, respectively, which is acceptable; the overhead of remote switching and row remapping is, on average 0.9\%, which is negligible.

Figure~\ref{eva(K-O)}(B) compares on-chip storage demand. That of Task Queues in the Omega-Network is in blue; the buffers for remote switching + row remapping are in red; the others are in green. As shown in Figure (B), the overall storage demands of AWB-GCN with Design(D) are even lower than the baseline. This is largely due to dramatically reduced per-PE Task Queue size under more balanced workloads. With much more balanced workload distributions in Design(D), the congestion and backpressure in Omega-Network are significantly relieved, making the TQs narrower and shallower.

Finally, Figure~\ref{autotuning} shows the utilization improvement due to iterative workload autotuning. Rebalancing can be accomplished within 10 iterations. This means that most of the iterations can benefit from operating under the converged optimal strategy. Note that the utilization of \emph{Nell} has a sharp improvement in round 3 due to effective evil row remapping.

\begin{table*}[h]
\scriptsize
\centering
\caption{Comparison with CPU, GPU and Baseline processing Standard\_networks. OoM: Out of Memory.}
\begin{tabular}{|c|c|c|c|c|c|c|}
\hline
\textbf{Platform}        & \textbf{Standard\_networks}             & \textbf{Cora}       & \textbf{CiteSeer} & \textbf{Pubmed}     & \textbf{Nell}     & \textbf{Reddit}      \\ \hline
Intel Xeon E5-2680 (PyG) & Latency (ms) {[}speedup{]}   & 2.51 {[}1$\times${]}       & 3.66 {[}1$\times${]}     & 13.97 {[}1$\times${]}      & 2.28E3 {[}1$\times${]}   & 2.94E5 {[}1$\times${]}      \\ \cline{2-7} 
Freq: 2.5GHz             & Energy efficiency (graph/kJ) & 6.68E3              & 3.88E3            & 1.03E3              & 6.99              & 5.43E-2              \\ \hline
NVIDIA RTX8000 (PyG)     & Latency (ms) {[}speedup{]}   & 0.69 {[}3.6$\times${]}     & 0.68 {[}5.4$\times${]}   & 0.69 {[}20.2$\times${]}    & 90.50 {[}25.2$\times${]} & OoM                   \\ \cline{2-7} 
Freq: 1395MHz            & Energy efficiency (graph/kJ) & 1.06E4              & 1.29E4            & 1.11E4              & 89.06             & OoM                   \\ \hline
SCNN (Cartesian product, 330MHz)     & Latency (ms) {[}speedup{]}   & 1.6E-2 {[}158$\times${]}     & 2.6E-2 {[}142$\times${]}   &7.2E-2 {[}195$\times${]}    & 31.47 {[}72.4$\times${]} &  73.22 {[}4015$\times${]}                  \\ \hline
Baseline Intel D5005 FPGA            & Latency (ms) {[}speedup{]}   & 1.3E-2 {[}191.8$\times${]} & 9.0E-3 {[}406$\times${]} & 6.7E-2 {[}208$\times${]} & 30.09 {[}75.8$\times${]} & 47.4 {[}6209$\times${]} \\ \cline{2-7} 
Freq: 330MHz             & Energy efficiency (graph/kJ) & 6.86E5              & 9.75E5            & 1.22E5              & 3.28E2            & 1.83E2               \\ \hline
AWB-GCN Intel D5005 FPGA             & Latency (ms) {[}speedup{]}   & 2.3E-3 {[}1063$\times${]}  & 4.0E-3 {[}913$\times${]} & 3E-2 {[}466$\times${]}   & 1.6 {[}1425$\times${]} & 31.81 {[}9242$\times${]}   \\ \cline{2-7} 
Freq: 330MHz             & Energy efficiency (graph/kJ) & 3.08E6              & 1.93E6            & 2.48E5              & 4.12E3            & 2.09E2               \\ \hline
\end{tabular}
\label{Table:cross_plat}
\end{table*}

\begin{table*}[h]
\scriptsize
\centering
\caption{Comparison with the prior art, HyGCN, processing HyGCN\_networks customized in HyGCN paper\cite{yan2020hygcn}.}
\begin{tabular}{|c|c|c|c|c|c|c|}
\hline
\textbf{Platform}        & \textbf{HyGCN\_networks}             & \textbf{Cora}       & \textbf{CiteSeer}   & \textbf{Pubmed}   & \textbf{Nell}      & \textbf{Reddit}      \\ \hline
Intel Xeon E5-2680 (PyG) & Latency (ms) {[}speedup{]}   & 13.07 {[}1$\times${]}      & 15.73 {[}1$\times${]}      & 2.19E2 {[}1$\times${]}   & 3.17E3 {[}1$\times${]}    & 8.05E5 {[}1$\times${]}      \\ \cline{2-7} 
Freq: 2.5GHz             & Energy efficiency (graph/kJ) & 1.23E3              & 9.36E2              & 70.61             & 4.59               & 0.02                 \\ \hline
NVIDIA RTX8000 (PyG)     & Latency (ms) {[}speedup{]}   & 0.69 {[}18.9$\times${]}    & 0.69 {[}22.8$\times${]}    & 1.31 {[}167$\times${]} & 100.18 {[}31.7$\times${]} & OoM                   \\ \cline{2-7} 
Freq: 1395MHz            & Energy efficiency (graph/kJ) & 1.16E4              & 1.09E4              & 6.46E3            & 79.81              & OoM                   \\ \hline
HyGCN TSMC 12 nm       & Latency (ms) {[}speedup{]}   & 2.1E-2 {[}627$\times${]} & 0.30 {[}52.1$\times${]}    & 0.64 {[}341.5$\times${]} & NA                 & 2.89E2 {[}2787$\times${]} \\ \cline{2-7} 
Freq: 1GHz               & Energy efficiency (graph/kJ) & 7.16E6              & 4.94E5              & 2.33E5            & NA                 & 5.17E2               \\ \hline
AWB-GCN Intel D5005 FPGA             & Latency (ms) {[}speedup{]}   & 1.7E-2 {[}768$\times${]} & 2.9E-2 {[}548$\times${]} & 0.23 {[}948$\times${]} & 3.25 {[}978$\times${]} & 49.7 {[}16197$\times${]}    \\ \cline{2-7} 
Freq: 330MHz             & Energy efficiency (graph/kJ) & 4.39E5              & 2.71E5              & 3.17E4            & 2.28E3             & 1.45E2               \\ \hline
\end{tabular}
\label{Table:cross_plat1}
\end{table*}





\subsection{Scalability of AWB-GCN}

We evaluate the scalability of AWB-GCN by running GCN inference of the five datasets on the baseline as well as Designs (B) and (D) of AWB-GCN and varying the number of PEs from 512, 1024, 2048 to 4096. In Figure~\ref{eva2}, the bars represent the performance speedup comparing with the baseline design with 512 PEs. The lines represent average PE utilizations. 

As shown in Figure~\ref{eva2}, the PE utilization of the baseline design drops dramatically with increasing number of PEs. This is because more PEs means fewer rows per PE, highlighting the imbalance among PEs: they have fewer opportunities to absorb inter-row imbalance. Due to the dropping PE utilization, the performance speedup shows poor scalability. For AWB-GCN with only distribution smoothing, PE utilization also drops but more slowly than baseline. \emph{Nell} is an outlier as the utilization of baseline with 512 PEs is too low to drop. In contrast, the PE utilization of the complete version of AWB-GCN, Design(D), is high and stable. The performance scales almost linearly with increasing number of PEs. 



\subsection{Cross-platform Comparison}


We evaluate five scenarios: (i) AWB-GCN Design-(D) with 4096 PEs, (ii) PyG-based implementation on Intel Xeon E5-2680v3 CPU (PyG-CPU), (iii) PyG-based implementation on a NVIDIA RTX 8000 GPU (PyG-GPU), (iv) 4096-PE baseline AWB-GCN without workload rebalancing, and (v) SCNN \cite{parashar2017scnn} reproduction with 4096 multipliers (we build a system-C-based cycle-accurate simulator for SCNN). We use the five datasets: Cora, Citeseer, Pubmed, Nell, and Reddit for the evaluation. The GCN model configuration follows the original GCN algorithm papers \cite{kipf2016semi,zhuang2018dual,chen2017stochastic}. We label these GCNs ``\emph{Standard\_networks}''.

As shown in Table~\ref{Table:cross_plat}, despite running at a relatively low frequency, AWB-GCN achieves, on average, speedups of $2622\times$ and $136\times$ over the well-optimized PyG implementations on high-end CPUs and GPUs. It achieves a speedup of $6.1\times$ over the baseline design without workload rebalancing. For the Nell dataset, the speedup over the baseline is 18.8$\times$, demonstrating in particular the impact of workload balancing. Reddit fails on GPU due to out-of-memory. AWB-GCN achieves from 2.3$\times$ to 19.7$\times$ speedup compared with SCNN. SCNN is inefficient when working with GCNs because it uses Cartesian Product-based SpMM which requires massive and highly irregular reduction of intermediate results, especially when the matrices are very big, sparse, and follow power-law. SCNN also requires very high off-chip bandwidth for the reduction. In our evaluation, we assume SCNN is equipped with a High Bandwidth Memory (HBM) which provides sufficient off-chip bandwidth. If DRAMs are used, the performance of SCNN would be even lower.




The tremendous speedups over PyG-CPU and PyG-GPU originate from AWB-GCN's dedicated architecture which uses features not available on general-purpose CPUs and GPUs: (a) the dynamic autotuning techniques ensure balanced workload and high PE utilization (Section 4); (b) all the SpMM kernels of the GCN layers are deeply pipelined, leading to reduced on-chip storage demand (Figure~\ref{kernel_parallelism}); (c) inter-layer data forwarding and matrix blocking (with column-wise-product sub-SpMM execution Figure~\ref{blocking_matrix}) improve data reuse and guarantee that off-chip memory accesses are to consecutive addresses. 

We also compare AWB-GCN with existing GCN accelerators. Prior to this work, to the best of our knowledge, the design proposed by Yan et al., HyGCN \cite{yan2020hygcn}, is the only reported accelerator of GCN inference. However, HyGCN customizes the hidden layer of all GCN models to 128 channels, which is distinct from the original settings \cite{kipf2016semi,zhuang2018dual,chen2017stochastic}. We refer to the HyGCN-customized models as \emph{HyGCN\_networks}. Also, the HyGCN report does not give absolute performance but, rather, relative speedups over a E5-2680v3 CPU. To compare AWB-GCN with HyGCN, we realize the \emph{HyGCN\_networks} on the same E5-2680v3 CPU, adopting the same software framework (PyG \cite{fey2019fast} -- HyGCN also uses PyG for the testing on CPU). The PyG-CPU result is thus a common baseline for comparing the relative speedups. Table~\ref{Table:cross_plat1} shows the results.


With the HyGCN\_networks, AWB-GCN achieves on average $3888\times$, $25.3\times$, and $5.1\times$ speedups over PyG-CPU, PyG-GPU, and the HyGCN design, respectively. The performance improvement is attributable, in part, to the features of AWB-GCN as discussed, and one additional reason: HyGCN scheduling is coarse-grained block-wise, while that of AWB-GCN is fine-grained element-wise. This avoids redundant data access and results in more benefit from a balanced workload.

As this design is implemented on an FPGA, it is difficult to compare energy efficiency to HyGCN, which is implemented as an ASIC. Comparing to ASIC, the reconfigurable routing switches on FPGA chip consume extra energy, making the energy efficiency of FPGA design lower (approximately 14$\times$ according to the numbers reported by Kuon \cite{4068926}). To compare the performance fairly, we limit the number of multipliers used and make it comparable to HyGCN. In particular, we use 4k 32-bit floating-point multipliers and HyGCN uses 4608 32-bit fixed-point multipliers. Floating-point multipliers also consume more energy than fixed-point ones.

\section{Related Work}

GNN studies use neural network algorithms to address problems in graph processing. The first GNN model was proposed by Gori et al. \cite{gori2005new}. In the past decade, work has continued on optimizing GNN algorithms exploring new neural network approaches 
\cite{dai2018learning,wu2020comprehensive,micheli2009neural,scarselli2008graph,you2018graphrnn,abu2018watch,gao2018large}. More recently, inspired by CNNs that achieve great success with euclidean data, GCNs are proposed for hidden feature extraction of non-euclidean data. 
In 2013, Bruna et al. \cite{bruna2013spectral} proposed the first GCNs for spectral graph theory; this was developed further in a number of variants \cite{henaff2015deep, defferrard2016convolutional,kipf2016semi}. GCNs are at the center of the research on neural-network-based graph processing \cite{yun2019graph}.

There have been many efforts on accelerating sparse CNNs \cite{kim2017novel,zhang2016cambricon,albericio2016cnvlutin,han2016eie,parashar2017scnn,kung2019packing,chen2016eyeriss,ding2017circnn}. We summarize them and explain why they fall short when applied to GCNs. Kung et al. condense the sparse parameter matrix through column grouping \cite{kung2019packing}. In case of conflict, only the most significant parameters are kept, others are discarded. Essentially, some accuracy is sacrificed for performance. Kim et al. \cite{kim2017novel} address the workload imbalance problem of sparse CNNs, but use information from design-time profiling and pre-scanning. Han et al. \cite{han2016eie} propose EIE, an SpMV accelerator that addresses imbalance with row-direction queuing. The design is not feasible in GCNs due to their large data size and power-law distribution. In EIE, weight matrices of SCNNs are distributively pre-stored on-chip in local buffers of PEs. This avoids off-chip non-zero accesses and online workload distribution, but is not possible for GCNs. Also, single-direction queuing fails to balance the workload of power-law matrices, which have serious imbalance on both directions. Zhang et al. \cite{zhang2016cambricon} propose Cambricon-S with efficient index matching to identify and multiplex non-zeros and feed them to massively parallel PEs. Again, these proposed architectures are not feasible for processing GCNs due to the ultra-low sparsity of power-law graphs which leads to highly scattered indices of neighboring elements. Given the adjacency matrix of Nell and a 1024-PE Cambricon-S, multiplexing enough non-zero pairs to feed all PEs per cycle would require 1024$\times$ 13699:1 multiplexers for single-precision floating point; this is not viable given likely chip technology.


Besides work on sparse CNNs, researchers also propose architectures for general SpMM. Zhuo and Prasanna \cite{zhuo2005sparse} present an SPMV design for FPGAs. Pal \cite{pal2018outerspace} proposes an outer-product-based SpMM architecture. This work focuses on reducing redundant memory accesses to non-zeros and does not essentially address the ultra-workload-imbalanced issue faced with GCNs. In their results, load-imbalances during the merge phase and the uneven data sharing patterns during the multiply phase lead to degraded speedup for the dataset with highly-unbalanced non-zero element distribution. 

SIGMA\cite{qin2020sigma} and ALRESCHA\cite{asgari2020alrescha} are recent high-performance architectures for SpMM and SpMV. We mainly discuss SIGMA, as SIGMA focuses on SpMM kernels and is equipped with more efficient optimizations for SpMM, while ALRESCHA has higher flexibility to support various kernels through switch reconfiguration. SIGMA uses an element-wise smart global controller to distribute every pair of non-zeros to the proper PEs dynamically through a Benes network. By doing so, PEs work with high utilization and the operations are evenly distributed among all multipliers so that workload imbalance is eliminated. SIGMA is highly efficient for general SpMMs, but needs some augmentation to work with GCNs. First, for a very large and sparse matrix, the type of bitmap compression format introduces significant overhead. 
Second, similar to Cambricon-S, the multiplexer required to get source/destination pairs would become very large, which limits the performance significantly.
Third, for the extremely large and sparse matrices common in GCN usage, the efficiency of the element-wise global controller decreases significantly when performing tasks such as matrix scanning and element filtering/counting which determines the number of Flex-DPEs.

To eliminate workload imbalance without using a global element-wise controller (as used in SIGMA), AWB-GCN uses auto-tuning-based rebalancing hardware, which is essentially also a ``controller'', to dynamically distribute tasks. In contrast to the controller used in SIGMA, AWB-GCN's is more coarse-grained and lighter weight. 
In particular, the distribution smoothing function is a {\it local} element-wise controller
which, similarly to SIGMA, distributes non-zero pairs to proper PEs. However, in contrast to SIGMA, in AWB-GCN the destination PEs must be local, meaning that they must within a few hops of the PE assigned in the initial mapping. Also, remote switching$+$row remapping is effectively a {\it global} row-wise controller
which can distribute tasks to any proper PEs without range limit, rather, with granularity of rows/fraction of rows instead of elements.
To make the proposed hybrid and light-weight controller handle workload imbalance as well as a global element-wise controller, we use auto-tuning. The proposed auto-tuning-based controller is designed especially for SpMMs with power-law matrices. For general SpMMs, a global element-wise controller can be more efficient.

Another active area of research is graph processing. Song et al.\cite{song2018graphr} and Zhang et al.\cite{zhang2018graphp} propose GraphR and GraphP, which are both based on Processing In Memory (PIM), to accelerate low-precision graph tasks. However, they do not support complex floating-point operations. Ham et al.\cite{ham2016graphicionado} propose Graphicionado, a vertex-centric acceleration framework for graph analytics applications.
It focuses on simple graph analysis applications. Ozdal et al.\cite{ozdal2016energy} propose a System-C based template for graph analytics applications. Ozdal's work and Graphicionado both use crossbars for data exchange which limits their scalability. None of these can directly support GCNs without significant modifications. 

Researchers also conduct software optimizations for SpMM on GPUs and general-purpose multicore CPUs \cite{Greathouse:2014,liu2014efficient,ashari2014fast,bell2008efficient,bell2009implementing}. These software solutions, however, do not meet the strict timing requirements of GCNs because of significant overhead in pre-scanning \cite{Greathouse:2014,liu2014efficient,ashari2014fast, yan2020hygcn} which is avoided in AWB-GCN. Also, adjacency matrices evolve at runtime, making offline processing even less useful.

\section{Conclusion}

In this paper, we propose AWB-GCN to accelerate GCN inference. To tackle the major performance issues derived from workload imbalance, we propose a hardware-based autotuning framework including three runtime workload rebalancing techniques: distribution smoothing, remote switching, and row remapping. The proposed rebalancing methods rely on hardware flexibility to realize performance autotuning with negligible area and delay overhead. This is the first accelerator design for GCNs that relies on hardware autotuning to achieve workload rebalancing for sparse matrix computations. We evaluate AWB-GCN using an Intel FPGA D5005 Accelerator Card with 5 widely used GCN datasets. Results show that AWB-GCN can achieve, on average, 3255$\times$, 80.3$\times$, and 5.1$\times$ speedups over high-end CPUs, GPUs, and other prior work respectively. Although FPGAs are used as a demonstration in this paper, the proposed architecture does not rely on any FPGA-specific features. And although AWB-GCN is designed for GCNs, it is generally efficient for GNNs whose major arithmetic primitives are also SpMMs, e.g., GraphSage \cite{hamilton2017inductive}, GINConv \cite{xu2018powerful}, and GTN \cite{yun2019graph}. 


\section*{Acknowledgements}
This work was supported, in part, by the NSF through Awards CCF-1618303, CCF-1919130, and CNS-1925504; by the NIH through Award R44GM128533; by grants from Microsoft and Red Hat; and by Xilinx and by Intel through donated FPGAs, tools, and IP. This research was also partially funded by Pacific Northwest National Laboratory's DMC-CFA project and DeepScience-HPC LDRD project. The evaluation platforms were supported by the U.S. DOE Office of Science, Office of Advanced Scientific Computing Research, under award 66150: ``CENATE - Center for Advanced Architecture Evaluation.'' The Pacific Northwest National Laboratory is operated by Battelle for the U.S. Department of Energy under contract DE-AC05-76RL01830. 

\bibliographystyle{ieeetr}
\bibliography{IEEEfull}


\end{document}